\documentclass[12pt]{article}
\usepackage{a4wide,epsfig,psfrag,amsmath,amssymb,cite,scalefnt}
\usepackage{color}

\newcommand{\gsim}{\;\rlap{\lower 3.5 pt \hbox{$\mathchar \sim$}} \raise 1pt
 \hbox {$>$}\;}
\newcommand{\lsim}{\;\rlap{\lower 3.5 pt \hbox{$\mathchar \sim$}} \raise 1pt
 \hbox {$<$}\;}

\newcommand{\mgluino}{m_{\tilde{g}}}
\newcommand{\mstopone}{m_{\tilde{t}_1}}
\newcommand{\mstoptwo}{m_{\tilde{t}_2}}
\newcommand{\msquark}{m_{\tilde{q}}}

\newcommand{\mtildeuR}{\tilde{m}_t}

\newcommand{\msusy}{m_{\rm susy}}

\newcommand{\xmus}{\frac{\mu_{\rm susy}}{\msusy}}
\newcommand{\lmMes}{l_{\varepsilon}}
\newcommand{\lmMsusy}{l_{\rm susy}}
\newcommand{\lmMt}{l_{\rm t}}
\newcommand{\lmMgl}{l_{\tilde{g}}}

\newcommand{\Salphapbeta}{s_{\alpha+\beta}}
\newcommand{\CalphadivSbeta}{\frac{c_\alpha}{s_\beta}}
\newcommand{\iCalphadivSbeta}{\frac{s_\beta}{c_\alpha}}
\newcommand{\Cfourthetat}{c_{4\theta_t}}
\newcommand{\Ctwothetat}{c_{2\theta_t}}
\newcommand{\TalphapdivTbeta}{\left(t_\alpha+\frac{1}{t_\beta}\right)}
\newcommand{\CSthetat}{c_{\theta_t} s_{\theta_t}}
\newcommand{\IEW}{\left(1 - \frac{8 s_W^2}{3}\right)}
\newcommand{\muSUSY}{\mu_{\rm susy}}

\newcommand{\softsusy}{SOFTSUSY}

\allowdisplaybreaks

\begin{document}

\title{\vskip-3cm{\baselineskip14pt
    \begin{flushleft}
      \normalsize SFB/CPP-12-57\\
      \normalsize TTP12-26\\
      \normalsize LPN12-087
  \end{flushleft}}
  \vskip1.5cm
  Supersymmetric next-to-next-to-leading order corrections to Higgs boson 
  production in gluon fusion
}

\author{
  Alexey Pak,
  Matthias Steinhauser,
  Nikolai Zerf
  \\[1em]
  {\small\it Institut f{\"u}r Theoretische Teilchenphysik}\\
  {\small\it Karlsruhe Institute of Technology (KIT)}\\
  {\small\it 76128 Karlsruhe, Germany}
}

\date{}

\maketitle

\thispagestyle{empty}

\begin{abstract}
  We compute the total cross section for the production of 
  a light CP even
  Higgs boson within the framework of supersymmetric QCD up to
  next-to-next-to-leading order. Technical subtleties in connection to the
  evaluation of three-loop Feynman integrals with many mass scales are
  discussed in detail and explicit results for the counterterms of the
  evanescent couplings are provided. The new results are applied to several
  phenomenological scenarios which are in accordance with the recent discovery
  at the LHC. In a large part of the still allowed parameter space the $K$
  factor of the supersymmetric theory is close to the one of the Standard
  Model. However, for the case where one of the top squarks is light, a
  deviation of more than 5\% in the next-to-next-to-leading order
  prediction of the cross section can
  be observed where at the same time the predicted Higgs boson mass has a
  value of about 125~GeV.  \medskip

  \noindent
  PACS numbers: 12.38.Bx 12.60.Jv 14.80.Da

\end{abstract}

\thispagestyle{empty}


\newpage


\section{Introduction}

Recently the experiments ATLAS and CMS at the Large Hadron Collider (LHC) at
CERN reported on the discovery of a new particle with a mass of about
125~GeV~\cite{ATLAS-Higgs,CMS-Higgs}. Although not all of its properties have been
determined yet the new particle seems to be in accordance with a Higgs boson,
maybe the one of the Standard Model (SM), maybe one corresponding to
extensions of the SM. In this paper we consider a supersymmetric extension of
the SM and report about next-to-next-to-leading order (NNLO) corrections to
the production cross section of the lightest CP even Higgs boson in this
model. One of the aims is to quantify the size of the supersymmetric
corrections.  First results in this direction have been reported in
Ref.~\cite{Pak:2010cu}. We will significantly extend these considerations 
and furthermore provide details on the calculation.

The radiative corrections both for inclusive and exclusive processes are
summarized in the reports~\cite{Dittmaier:2011ti} and~\cite{Dittmaier:2012vm},
respectively.  The numerical results for the SM Higgs boson production
cross section provided in these papers are based on calculations to the
gluon fusion process which have been performed up to next-to-leading
order (NLO)~\cite{Georgi:1977gs,Djouadi:1991tka,Dawson:1990zj,Spira:1995rr}
for arbitrary values of the top quark and Higgs boson mass and to
NNLO in
Refs.~\cite{Harlander:2002wh,Anastasiou:2002yz,Ravindran:2003um} adopting the
infinite top quark mass
approximation. NNLO results for finite top quark masses
have been obtained in
Ref.~\cite{Harlander:2009bw,Pak:2009bx,Harlander:2009mq,Pak:2009dg,Harlander:2010my,Pak:2011hs}.

Within the SM the bottom quark contributions are small and NLO
terms~\cite{Spira:1995rr} are sufficient considering the size of the
corrections induced by the top quark Yukawa coupling.  This is different in
the case of supersymmetry (SUSY) where the factor $m_b$ is accompanied by
$\tan\beta$, the ratio of the vacuum expectation values of the two Higgs
doublets, and can thus lead to large contributions to the cross section for
the production of the lighter CP even Higgs boson. Explicit analytical NLO
results can be found in Refs.~\cite{Degrassi:2010eu,Harlander:2010wr} (see
also Ref.~\cite{Anastasiou:2008rm} for numerical results).  The contribution
from the top quark/top squark sector is available already since a few
years. Within the effective-theory framework, which constitutes an excellent
approximation for Higgs boson masses below about 200~GeV, the production cross
section has first been computed in
Refs.~\cite{Harlander:2003bb,Harlander:2004tp} and has been confirmed
analytically~\cite{Degrassi:2008zj} and numerically~\cite{Anastasiou:2008rm}
(see also Ref.~\cite{Muhlleitner:2010nm}). In
Refs.~\cite{Muhlleitner:2006wx,Bonciani:2007ex} the squark loop
contributions to Higgs boson production in the Minimal Supersymmetric
  Standard Model (MSSM) have been computed without
assuming any mass hierarchy.

The first supersymmetric QCD
corrections at NNLO have been computed in Ref.~\cite{Pak:2010cu} (see also
Refs.~\cite{Harlander:2003kf,Harlander:2004tp} for estimates prior to this
work) for a degenerate mass spectrum. In this work these considerations are
extended to allow for a wide range of phenomenologically interesting
scenarios. The three-loop results for the matching coefficients which are
presented in this paper have been confirmed by an independent calculation in
Ref.~\cite{Kurz:2012ff}. 
Note that the calculations 
in the paper are performed in the framework of the MSSM,
however, they are in principle not limited to the minimal version and
  could, e.g.,  also be applied to the so-called 
  Next-to-Minimal Supersymmetric Standard Model (NMSSM) where an additional
  singlet Higgs field is added as compared to the MSSM.

In Ref.~\cite{Pak:2010cu} all supersymmetric masses have been
identified. Furthermore, some of
the technical difficulties in connection to the interplay of dimensional
reduction (DRED) and
supersymmetry have been addressed in this reference. In this context it has
been very useful to evaluate the QCD corrections within DRED and check that the
proper renormalization of the evanescent couplings leads to the known analytical
expressions after re-expressing the strong DRED coupling by its
counterpart in dimensional regularization (DREG). 
In this paper we build on the developed formalism and
evaluate the Higgs-gluon coupling for various hierarchies which are of
phenomenological relevance. In particular, we provide the counterterms of the
evanescent couplings which could be relevant also for other calculations.

The outline of the paper is as follows: in the next section we briefly review
the formalism and discuss the subtleties at NLO (where an exact calculation
has been performed) and NNLO. Moreover the quality of the approximations
applied at NNLO are tested at the NLO level. We discuss at length the
renormalization procedure. Numerical results for the Higgs-gluon coupling are
also presented in this Section.  Afterwards we compute in
Section~\ref{sec::num} predictions for the total Higgs boson production cross
section and consider different scenarios: $m_h^{\rm max}$ which maximizes the
prediction of the lighter CP even MSSM Higgs boson and a scenario where one of
the top squarks is light whereas the reminder of the mass spectrum is heavy.
Our conclusions are presented in Section~\ref{sec::concl}.


\section{Higgs-gluon coupling in the MSSM}


\subsection{Effective theory}

The theoretical framework needed for the computation of NNLO corrections to
the production of a light CP even Higgs boson has already been described in
Ref.~\cite{Pak:2010cu}. For completeness we repeat the most important issues
and formulae.  Starting point is the effective five-flavour Lagrange density
\begin{eqnarray}
  {\cal L}_{Y,\rm eff} &=& -\frac{\phi^0}{v^0} C_1^0 {\cal O}_1^0 
  + {\cal L}_{QCD}^{(5)}
  \,,
  \label{eq::leff}
\end{eqnarray}
where ${\cal L}_{QCD}^{(5)}$ is the usual QCD part with five massless quarks.
$\phi^0$ denotes the light CP-even Higgs boson of the MSSM, $v^0$ is given by
$\sqrt{(v^0_1)^2 + (v^0_2)^2}$ where $v_1$ and $v_2$ are the vacuum
expectation values of the two Higgs doublets, and $G_{\mu\nu}^0$ is the gluonic
field strength tensor constructed from fields and couplings already present in
${\cal L}_{QCD}^{(5)}$. The superscript ``0'' denotes bare quantities. Note
that the renormalization of $\phi^0/v^0$ is of higher order in the
electromagnetic coupling constant.  In Eq.~(\ref{eq::leff}) the operator
${\cal O}_1^0$ is given by
\begin{eqnarray}
  {\cal O}_1^0 &=& \frac{1}{4} G_{\mu\nu}^0 G^{0,\mu\nu}\,.
\end{eqnarray}
It contains the coupling of the Higgs boson to two, three and four gluons.
Heavy degrees of freedom only contribute to the coefficient function
$C_1^0$. In the SM this concerns only the top quark whereas 
for supersymmetric QCD also squarks and gluinos contribute.

\begin{figure}[t]
  \centering
  \includegraphics[width=\linewidth]{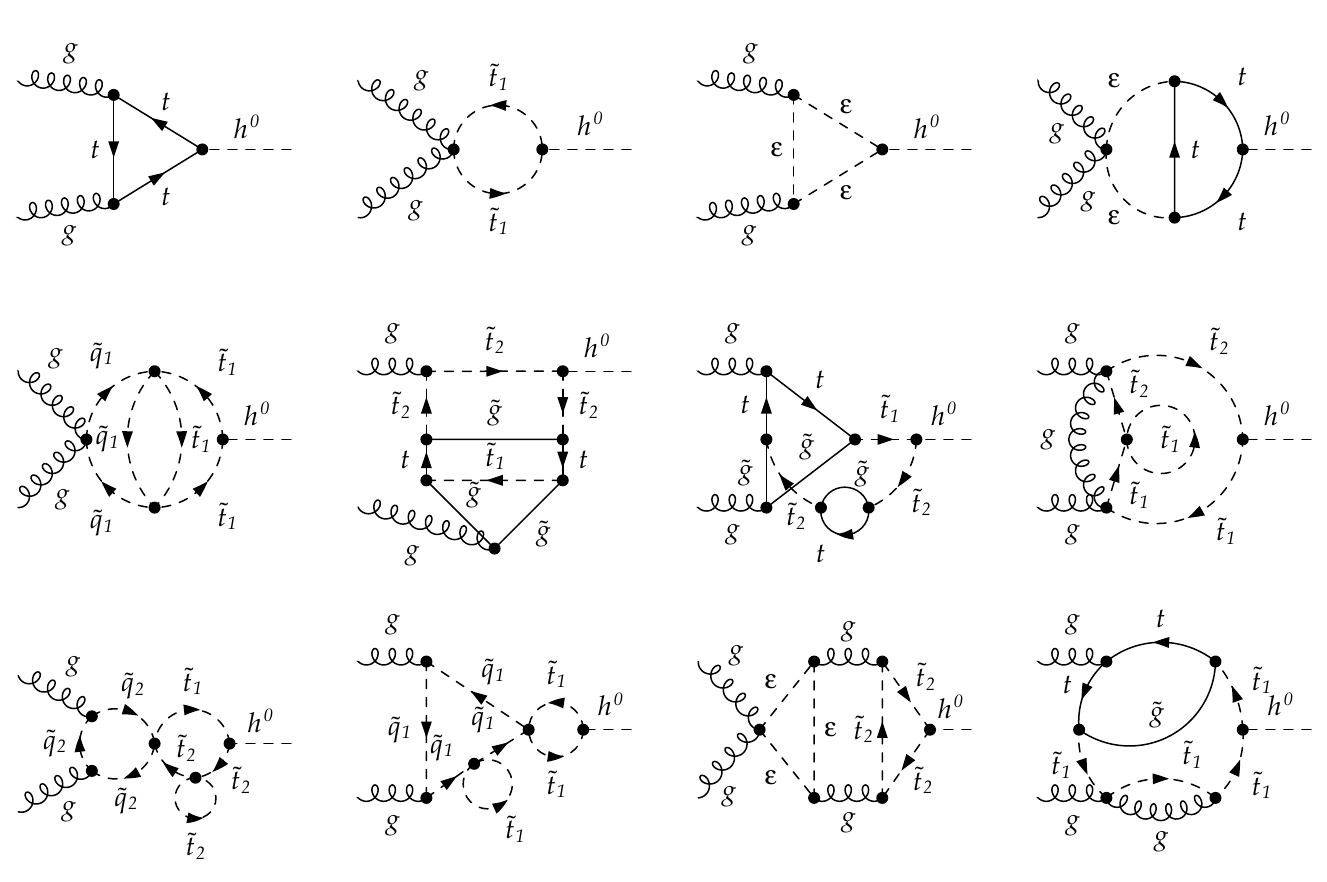}
  \caption[]{\label{fig::diags1}Sample diagrams contributing to $C_1$ at one,
    two and three loops originating from the coupling of the Higgs boson
      to top quarks and top squarks.  The symbols $t$, $\tilde{t}_i$, $g$,
    $\tilde{g}$, $h$ and $\varepsilon$ denote top quarks, top squarks, gluons,
    gluinos, Higgs bosons and $\varepsilon$ scalars, respectively.}
\end{figure}

\begin{figure}[t]
  \centering
  \includegraphics[width=\linewidth]{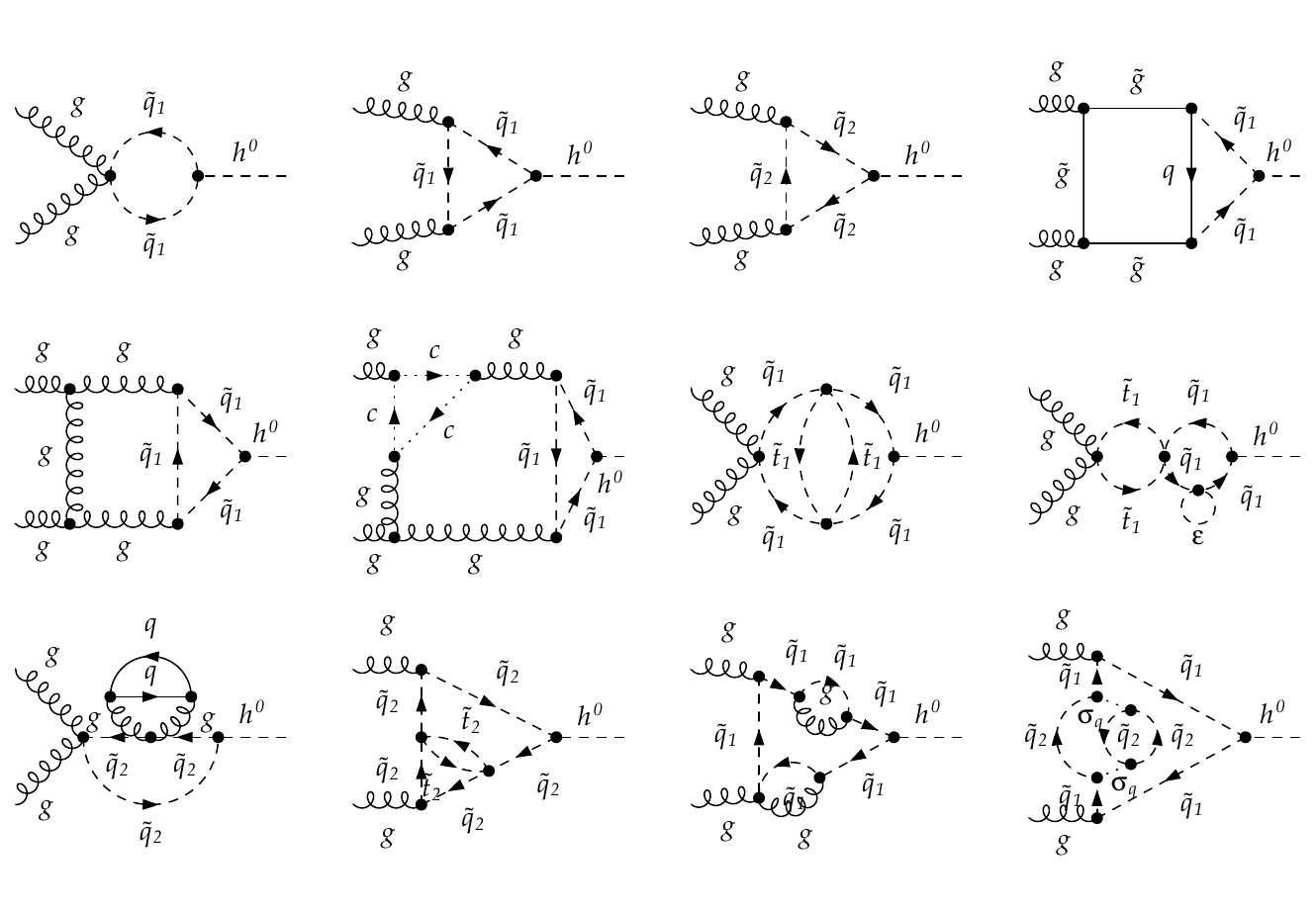}
  \caption[]{\label{fig::diags2}Sample diagrams contributing to $C_1$ at one,
    two and three loops originating from the coupling of the Higgs boson
      to the squarks corresponding to the light quarks.  The symbols $q$,
    $\tilde{q}_i$, $g$, $\tilde{g}$, $c$, $h$ and $\varepsilon$ denote light
    quarks, the corresponding scalar super partners, gluons, gluinos, ghosts,
    Higgs bosons and $\varepsilon$ scalars, respectively.  $\sigma_q$ is an
    auxiliary particle used for the implementation of the four-squark vertex.}
\end{figure}

From the technical point of view $C_1^0$ is computed from vertex diagrams
involving two gluons and a Higgs boson.  In the practical calculation it is
convenient to subdivide the contributing Feynman diagrams into two
classes. The first one contains all contributions where the external Higgs
boson couples to a top quark or top squark (see Fig.~\ref{fig::diags1} for
sample diagrams) whereas all diagrams with a coupling of the Higgs boson to a
super partner of the five lighter quark flavours constitute the second class
(see Fig.~\ref{fig::diags2}).  Note that the diagrams of the latter class lead
to corrections to $C_1^0$ which are suppressed by $M_Z^2/m_{\tilde{q}}^2$
whereas the Higgs-top-squark coupling generates contributions proportional to
$1$, $m_t^2/m_{\tilde{t}}^2$, $m_t\muSUSY/m_{\tilde{t}}^2$ and
  $M_Z^2/m_{\tilde{t}}^2$ (where $\muSUSY$ is the Higgs-Higgsino bilinear
  coupling from the super potential). Although the relative suppression is
formally only $M_Z^2/m_t^2$ all $M_Z^2/m_{\tilde{q}}^2$ terms are numerically
much smaller than the leading contribution.  In our calculation we consider
all squarks except the top squark to be degenerate and we furthermore assume
that the corresponding quarks are massless. Thus the squark contributions from
the first two families cancel and only the contribution from bottom squarks
remains. In the following we nevertheless use the notion ``squark contribution''
for diagrams from Fig.~\ref{fig::diags2}.

In this paper we compute all contributions from the top quark/top squark and 
light quark/squark sector in the limit of vanishing light quark masses up to
three loops. For the corresponding Feynman rules we refer to the Appendices of
Refs.~\cite{Harlander:2004tp} and~\cite{Harlander:2010wr}. Note that the
couplings to the super partners of the light quarks has been implemented by
averaging over the up, down, strange, charm and bottom squark contribution.
Since for large values of $\tan\beta$ bottom quark diagrams might
give large contributions they will be included in our numerical prediction up
to NLO~\cite{Degrassi:2010eu,Harlander:2010wr} as we will describe in
Section~\ref{sub::form}. 

Since $C_1^0$ constitutes the effective Higgs-gluon coupling it is sufficient
to evaluate the corresponding Feynman integrals in the limit of vanishing
external momenta. This is conveniently achieved by applying a projector (see,
e.g., Eqs.~(11) and~(12) in Ref.~\cite{Pak:2010cu}) on one of the two tensor
structures and by performing a naive Taylor expansion afterwards. As a result
one ends up with vacuum integrals involving several mass scales in the case of
supersymmetry. Up to two-loop order such integrals can be computed exactly,
the corresponding algorithm and necessary formulae have been provided in
Ref.~\cite{Davydychev:1992mt} almost 20~years ago.  Since an analogue
algorithm at three-loop order is not available to date one has to rely on
special cases combined with expansions in order to construct approximations to
the exact formulae.  A general outline of such an approach has been discussed
in Ref.~\cite{Kant:2010tf} where three-loop corrections to the lightest Higgs
boson mass have been computed.  In this paper we refine the approximation
method which is described in Subsection~\ref{sub::appr}. However, we first
discuss in the next subsection the renormalization procedure.


\subsection{Renormalization and evanescent couplings}

Besides the technical difficulties, which are described in
Subsection~\ref{sub::appr}, also several field theoretic
challenges have to be solved. They are mainly connected to the fact that for
the calculation of $C_1^0$ it is convenient to use
DRED~\cite{Siegel:1979wq,Harlander:2006xq,Harlander:2006rj,Jack:2007ni}
whereas the calculations within the effective theory of
Eq.~(\ref{eq::leff}) are performed with the help of 
DREG. The LO calculation is not affected by
this issue. Also at NLO it is possible to proceed in a naive way. 
At NNLO, however,
it is necessary to properly renormalize all occurring parameters and to use
the relations between their DRED and DREG definition.  In
Ref.~\cite{Pak:2010cu} the procedure has been described for a degenerate SUSY
spectrum, the modifications necessary to cover the most general case are
straightforward and briefly touched in the following.

We renormalize all parameters in the $\overline{\rm DR}$ scheme except
$M_\varepsilon$ and the coupling of the Higgs boson to $\varepsilon$
scalars. $M_\varepsilon$ is renormalized on-shell with the condition
$M_\varepsilon=0$. Note that we take this limit only in the final result
since in intermediate steps $1/M_\varepsilon$ terms occur which have to be
canceled by proper counterterms.
The two-loop $\overline{\rm DR}$ counterterms for
$\alpha_s$ and $m_t$ are independent of all occurring scales and can be found
in Refs.~\cite{Kant:2010tf,Hermann:2011ha} (see also references therein). 
Due
to the quadratic divergence the 
renormalization constants for the squark masses and the mixing angle
$\theta_t$, which we also need up to two loops, depend on mass ratios which is
known exactly, even up to three loops~\cite{Hermann:2011ha}.
The gluino mass counterterm is only needed to one-loop order.  As
far as wave function renormalization of external particles
is concerned only the one for the
external gluons contributes. Actually, for our calculation only those 
contributions to the gluon two-point function have to be considered which
contain heavy particles. This is conveniently formulated in terms 
of a bare decoupling
constant of the gluon field~\cite{Chetyrkin:1997un} which is usually denoted
by $\zeta_3^0$. 
Note that in our case $\zeta_3^0$ has to accomplish both the
decoupling of all supersymmetric particles and the top quark and the
transition from DRED to DREG. For the degenerate mass case results can be
found in Ref.~\cite{Pak:2010cu}. The analytical expressions needed for the
calculations performed in this paper can be found in 
electronic form in~\cite{progdata}. Note that $C_1^0$ is multiplicatively
renormalized with $1/\zeta_3^0$.

A non-standard part of the renormalization procedure concerns the so-called
evanescent couplings which emerge through radiative corrections. In our case
there are two operators which define the following
Lagrange density (see also discussion in Ref.~\cite{Anastasiou:2008rm})
\begin{eqnarray}
  {\cal L}_\varepsilon &=& 
  -\frac{1}{2}\left(M_\varepsilon^0\right)^2 
  \varepsilon^{0,a}_\sigma\varepsilon^{0,a}_\sigma
  - \frac{\phi^0}{v^0} \left(\Lambda_\varepsilon^0\right)^2
  \varepsilon^{0,a}_\sigma\varepsilon^{0,a}_\sigma 
  \,,
  \label{eq::Lep}
\end{eqnarray}
where $\varepsilon^{0,a}$ denotes the bare $\varepsilon$ scalar field and the
dimensionful quantity $\Lambda_\varepsilon^0$ mediates the coupling of the
Higgs boson to $\varepsilon$ scalars. The on-shell $\varepsilon$ scalar mass
counterterm is needed to one-loop accuracy and is given
by (see, e.g., Ref.~\cite{Bednyakov:2002sf})
\begin{eqnarray}
  \left( M_\varepsilon^0 \right)^2 &=&  M_\varepsilon^2 + \delta M_\varepsilon^2
  \nonumber\\
  \delta M_\varepsilon^2 &=& \frac{\alpha_s^{\rm (SQCD)}}{\pi}\Bigg\{
  \Bigg[
   \frac{1}{4}\mstopone^2\left(\frac{\mu^2}{\mstopone^2}\right)^\epsilon
  +\frac{1}{4}\mstoptwo^2\left(\frac{\mu^2}{\mstoptwo^2}\right)^\epsilon
  +\frac{n_l}{2}\msquark^2\left(\frac{\mu^2}{\msquark^2}\right)^\epsilon
  -\frac{3}{2}\mgluino^2\left(\frac{\mu^2}{\mgluino^2}\right)^\epsilon
  \nonumber\\&&\mbox{}
  -\frac{1}{2}m_t^2\left(\frac{\mu^2}{m_t^2}\right)^\epsilon
    \Bigg]
    \Bigg[\frac{1}{\epsilon} + 1 + \epsilon\left(1+\frac{\pi^2}{12}\right)\Bigg]
  \nonumber\\&&\mbox{}
  + M_\varepsilon^2\Bigg[
  - \frac{2}{\epsilon} 
  - \frac{9}{2} 
  + \frac{3\lmMgl}{4} 
  + \frac{\lmMt}{4} 
  - 3\lmMes
  + \epsilon\left(-\frac{21}{2} -\frac{\pi^2}{6} 
    + \frac{3\lmMgl^2}{8} 
    + \frac{\lmMt^2}{8}
    - \frac{9\lmMes}{2} 
    - \frac{3\lmMes^2}{2} 
  \right)
  \nonumber\\&&\mbox{}
  + n_l\Bigg(
    \frac{1}{4\epsilon} 
  + \frac{1}{2} 
  + \frac{\lmMes}{4}
  + \epsilon\left(
    1 - \frac{7\pi^2}{48} 
    + \frac{\lmMes}{2} 
    + \frac{\lmMes^2}{8} 
  \right)
  \Bigg)
  \Bigg]
  + {\cal O}\left(M_\varepsilon^4\right)
  + {\cal O}\left(\epsilon^2\right)
  \,,
\end{eqnarray}
with $l_x=\ln(\mu_R^2/m_x^2)$ and $l_{\varepsilon} =
\ln(\mu_R^2/M_\varepsilon^2)$ where $\mu_R$ is the renormalization scale.
$n_l=5$ counts the number of massless quarks.  Note that for our calculation
the terms of ${\cal O}(M_\varepsilon^2)$ have to be kept until the inverse
powers in the $\varepsilon$ scalar mass are canceled (cf. discussion below).

\begin{figure}[t]
  \centering
  \includegraphics[width=\linewidth]{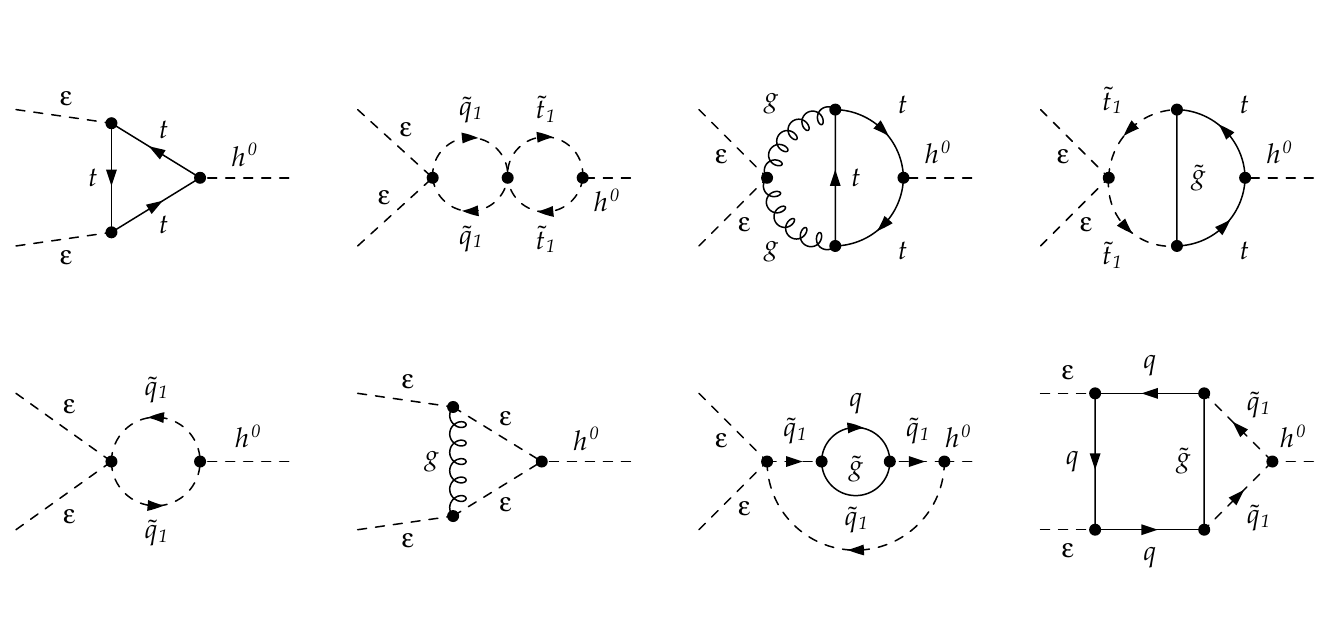}
  \caption[]{\label{fig::diags3}One- and two-loop sample diagrams contributing
    to $\delta\Lambda_\varepsilon$. The notation is adopted from
    Figs.~\ref{fig::diags1} and ~\ref{fig::diags2}.}
\end{figure}

The counterterm for $\Lambda_\varepsilon^0$ is defined through
\begin{eqnarray}
  \left(\Lambda_\varepsilon^0\right)^2
  &=&
  \delta\Lambda_{\varepsilon}^2 + \Lambda_\varepsilon^2
\end{eqnarray}
and is needed to two-loop accuracy. Sample Feynman diagrams contributing to
$\delta\Lambda_{\varepsilon}$ are shown in Fig.~\ref{fig::diags3}.  We apply
the renormalization condition $\Lambda_\varepsilon=0$.  The analytical
expressions corresponding to the three hierarchies are quite lengthy --- in
particular for those cases where $\theta_t\not=0$ --- and can be found on the
webpage~\cite{progdata}. To illustrate the structure of the result we present
in the following $\delta\Lambda_{\varepsilon}^2$ for the hierarchy (h1) which
is defined in Eq.~(\ref{eq::hier}). It is convenient to split the result 
into  a
contribution from the top quark/top squark sector and from the light
quark/squark sector. Thus the leading terms read
\begin{eqnarray}
  \left(\delta\Lambda_{\varepsilon}^{(\rm h1)}\right)^2 
  &=& 
  \left(\delta\Lambda_{\varepsilon,\tilde{t}}^{(\rm h1)}\right)^2 
  +
  \left(\delta\Lambda_{\varepsilon,\tilde{q}}^{(\rm h1)}\right)^2 
  \,,
  \nonumber\\
  \left(\delta\Lambda_{\varepsilon,\tilde{q}}^{(\rm h1)}\right)^2
  &=&
  -2 M_Z^2 \Salphapbeta\Bigg\{
  \frac{\alpha_s^{\rm (SQCD)}}{\pi}
  \Bigg[ \frac{1}{8\epsilon} + \frac{\lmMsusy}{8} 
  + \epsilon\left(\frac{\lmMsusy^2}{16} + \frac{\pi^2}{96}\right)
  \Bigg]
  \nonumber\\&&\mbox{}
  + \left(\frac{\alpha_s^{\rm (SQCD)}}{\pi}\right)^2\!\Bigg[
  \frac{-8+n_l}{32\epsilon^2}
  + \frac{6 + (-8+n_l)\lmMsusy}{32\epsilon}
  - \frac{9}{16} 
  - \frac{\pi^2}{48} 
  + \frac{5\lmMsusy}{24} 
  - \frac{\lmMsusy^2}{8} 
  \nonumber\\&&\mbox{}
  + n_l\left(
    \frac{\lmMsusy^2}{64} 
    + \frac{\pi^2}{384}
  \right)
  + \frac{M_\varepsilon^2}{\msusy^2}
  \left( \frac{5}{32} + \frac{7\lmMes}{192} - \frac{7\lmMsusy}{192} \right)
  \Bigg]
  \Bigg\}
  \,,
  \nonumber\\
  \left(\delta\Lambda_{\varepsilon,\tilde{t}}^{(\rm h1)}\right)^2
  &=&
-2m_t^2\frac{c_\alpha}{s_{\beta}} 
                            \Bigg\{
                                      \frac{\alpha_s^{(\text{SQCD})}}{\pi}
                                      \Bigg[
                                               \frac{1}{2}l_{\text{susy}} 
                                             -  \frac{1}{2}l_t 
                                             + \epsilon 
                                               \bigg(
                                                      \frac{1}{4}l_{\text{susy}}^2 
                                                     - \frac{1}{4}l_t^2
                                               \bigg)
                                       \Bigg]\nonumber\\
                                  &&\mbox{} + \left(\frac{\alpha_s^{(\text{SQCD})}}{\pi}\right)^2 \!\!
                                      \Bigg[
                                              \frac{1}{\epsilon}
                                               \bigg(
                                                      -l_{\text{susy}}
                                                      +l_t 
                                                      +n_l
                                                       \Big(
                                                              \frac{1}{8}l_{\text{susy}} 
                                                            - \frac{1}{8}l_t
                                                       \Big) 
                                               \bigg)
                                               + \frac{1}{3} 
                                             + \frac{23}{12} l_{\text{susy}} 
                                             - \frac{1}{8}l_{\text{susy}}^2 
                                               \nonumber\\ &&\mbox{}
                                             - \frac{23}{12} l_t 
                                             - \frac{1}{12}l_{\text{susy}} l_t 
                                             + \frac{5}{24} l_t^2 
                                             + n_l
                                               \bigg(
                                                       \frac{1}{16}l_{\text{susy}}^2 
                                                     - \frac{1}{16}l_t^2
                                               \bigg)
  + \frac{M_\varepsilon^2}{m_t^2}
  \Bigg( -\frac{7}{24} +\frac{l_t}{3} - \frac{\lmMes}{3} \Bigg)
                                               \nonumber\\   &&\mbox{}
  + \frac{M_\varepsilon^2}{\msusy^2}
  \Bigg( \frac{5}{8} 
  + \frac{7\lmMes}{48} - \frac{7\lmMsusy}{48} \Bigg)
                                             + \frac{m_t\muSUSY}{\msusy^2}
                                               \bigg(
                                                      t_{\alpha}
                                                     +\frac{1}{t_{\beta}}
                                               \bigg)
                                               \bigg(
                                                       \frac{11}{12} 
                                                     + \frac{1}{6}l_{\text{susy}} 
                                                     + \frac{1}{6}l_t
                                               \bigg)
\nonumber\\&&\mbox{}
                                               + \frac{m_t^2}{\msusy^2}
                                               \bigg(
                                                       -\frac{359}{216} 
                                                       +\frac{19}{72} l_{\text{susy}} 
                                                       -\frac{67}{72} l_t 
                                               \bigg) 
                                      \Bigg]
                            \Bigg\}
%
  \!-2 M_Z^2 \Salphapbeta\Bigg\{
  \frac{\alpha_s^{\rm (SQCD)}}{\pi}
  \Bigg[\! - \frac{1}{8\epsilon} - \frac{\lmMsusy}{8} 
  \nonumber\\&&\mbox{}
  - \epsilon\left(\frac{\lmMsusy^2}{16} + \frac{\pi^2}{96}\right)
  \Bigg]
  + \left(\frac{\alpha_s^{\rm (SQCD)}}{\pi}\right)^2\Bigg[
  \frac{8-n_l}{32\epsilon^2}
  + \frac{-6 - (-8+n_l)\lmMsusy}{32\epsilon}
  + \frac{9}{16} 
  \nonumber\\&&\mbox{}
  + \frac{\pi^2}{48} 
  - \frac{5\lmMsusy}{24} 
  + \frac{\lmMsusy^2}{8} 
  - n_l\left(
    \frac{\lmMsusy^2}{64} 
    + \frac{\pi^2}{384}
  \right)
  + \frac{M_\varepsilon^2}{\msusy^2}
  \left(-\frac{5}{32} - 
    \frac{7\lmMes}{192} + \frac{7\lmMsusy}{192} \right)
  \nonumber\\&&\mbox{}
  + \frac{m_t^2}{\msusy^2}\left(\frac{11}{108} + \frac{43\lmMsusy}{288} +
    \frac{5\lmMt}{288}
    \right)
  \Bigg]
  \Bigg\}
  + {\cal O}\left(\frac{m_t^3}{\msusy^3}\right)
  \,,
  \label{eq::deltalambda}
\end{eqnarray}
where $\msusy=\mstopone=\mstoptwo=\msquark=\mgluino$, 
$c_x=\cos(x)$, $s_x=\sin(x)$ and $t_x=\tan(x)$.
Here $\alpha$ is the
mixing angle between the weak and the mass eigenstates of the neutral
scalar Higgs bosons and $\tan\beta$ is the ratio of the vacuum expectation values
of the two Higgs doublets.

A closer look to Eq.~(\ref{eq::deltalambda}) shows that the terms proportional
to $M_Z^2$ exactly cancel in the sum of
$\left(\delta\Lambda_{\varepsilon,\tilde{t}}^{(\rm h1)}\right)^2$ and
$\left(\delta\Lambda_{\varepsilon,\tilde{q}}^{(\rm h1)}\right)^2$.  Thus for a
massless top quark there is no need to introduce a counterterm for
$\Lambda_\varepsilon$ since the corresponding contributions cancel within one
family. Note, however, that the $M_Z^2$ terms are needed to obtain separately
finite results for the top and light quark sector.  

In the supersymmetric limit the appearance of the evanescent coupling
$\delta\lambda_{\varepsilon}$ is forbidden by supersymmetry and thus there
will be no radiative corrections to a vanishing $h\varepsilon\varepsilon$
coupling. Both $\delta\Lambda_{\varepsilon,\tilde{t}}$ and
$\delta\Lambda_{\varepsilon,\tilde{q}}$ vanish independently.  Any non
vanishing radiative corrections to the $h\varepsilon\varepsilon$ coupling is
generated by the presence of soft SUSY breaking terms in the Lagrangian.

It is interesting to note that without introducing
$\delta\Lambda_{\varepsilon}$ the terms in $C_1$ proportional to $M_Z^2$ are
not finite in the limit $M_\varepsilon\to 0$. Rather they behave like
$M_Z^2/M_\varepsilon^2$ at two-loop order and even like
$M_Z^4/M_\varepsilon^4$ at three loops which is due to diagrams like the third
one in the third row of Fig.~\ref{fig::diags1}.  The inverse powers in
$M_\varepsilon$ only cancel after the contribution from
$\delta\Lambda_{\varepsilon}$ is added and thus the limit for vanishing
$\varepsilon$ scalar mass can be taken in the final expression. The
cancellation works independently for the top (Fig.~\ref{fig::diags1}) and
light quark sector (Fig.~\ref{fig::diags2}).  In Eq.~(\ref{eq::deltalambda})
we also list the contributions of order $M_\varepsilon^2/m_t^2$ and
$M_\varepsilon^2/\msusy^2$ which, however, do not contribute to the final
result.  In intermediate steps also $m_t^2/M_\varepsilon^2$ terms are present
which, however, cancel among the diagrams involving top quark and top squark
loops.

From this discussion it becomes clear that the $1/M_\varepsilon^n$ terms could
be avoided if a matching to an effective Lagrangian defined within DRED would
be performed. However, in that case it is likely that more operators have to
be considered in the effective Lagrangian which have not yet been
studied in the literature.  Alternatively the calculations could probably
also be performed in the limit where $M_\varepsilon$ is the largest
scale. Since the final result has to be independent of $M_\varepsilon$ one
could perform the limit $M_\varepsilon\to \infty$. As a consequence, the
inverse powers in $M_\varepsilon$ drop out by construction and the
positive powers have to cancel in the final result, in analogy to the
inverse powers in our approach.

The next step concerns the transformation of the $\overline{\rm DR}$ strong
coupling in the full theory ($\alpha_s^{\rm (SQCD)}$) to the $\overline{\rm
  MS}$ version in five-flavour QCD ($\alpha_s^{(5)}$). This is again achieved
with the help of the corresponding decoupling constant defined through
\begin{eqnarray}
  \alpha_s^{(5)} &=& \zeta_{\alpha_s} \alpha_s^{\rm (SQCD)}
  \,.
  \label{eq::zetaas}
\end{eqnarray}
Results for $\zeta_{\alpha_s}$ can again be found in Refs.~\cite{Pak:2010cu}
for the degenerate mass case. Exact results can be obtained from
the calculations performed in Ref.~\cite{Bauer:2008bj} but also
from~\cite{Kurz:2012ff} where an explicit result is provided.
Note that it is important to have at hand
the one-loop term of $\zeta_{\alpha_s}$ up
to order $\epsilon$ since at this step $C_1^0$ still contains poles
originating from the light degrees of freedom. They
are removed with the help of the renormalization constant
\begin{eqnarray}
  Z_{11} &=& \left(1 -
    \frac{\pi}{\alpha_s^{(5)}}\frac{\beta^{(5)}}{\epsilon}\right)^{-1}
  \nonumber\\
  &=& 1 
  + \frac{\alpha_s^{(5)}}{\pi}
  \frac{1}{\epsilon}
  \left(\frac{11}{4} - \frac{n_l}{6}\right)
  + \left(\frac{\alpha_s^{(5)}}{\pi}\right)^2
  \frac{1}{\epsilon}
  \left(\frac{51}{8} - \frac{19n_l}{24}\right)
  \,,
\end{eqnarray}
via
\begin{eqnarray}
  C_1 &=& \frac{1}{Z_{11}} C_1^0\,,
\end{eqnarray}
where $\beta^{(5)}= -(\alpha_s^{(5)}/\pi)^2\beta_0^{(5)} + \ldots$ is the
five-flavour QCD beta function with $\beta_0^{(n_l)}=11/4-n_l/6$ and in the
second line the perturbative expansion up to two loops is displayed.

\begin{table}
  \begin{center}
    \renewcommand{\arraystretch}{1.2}
  \begin{tabular}{c|l|c}
    \hline
    renormalization constant & explanation & loop order \\
    \hline
    $Z_{\alpha_s}$ & strong coupling constant & 2 \\
    $Z_{m_{\tilde{t}_i}}$ & top squark mass, $i=1,2$ & 2 \\
    $Z_{m_{\tilde{q}}}$ & squark mass & 2 \\
    $Z_{m_{\tilde{g}}}$ & gluino mass & 1 \\
    $\delta M_\varepsilon^2$ & $\varepsilon$ scalar mass & 1 \\
    $\delta\Lambda_\varepsilon^2$ & $h\varepsilon\varepsilon$ coupling & 2 \\
    $Z_{11}$ & operator renormalization & 2 \\
    \hline
    decoupling constant & explanation & loop order \\
    \hline
    $\zeta_3^0$    & decouples heavy particles form gluon field & 2 \\
    $\zeta_{\alpha_s}$      & relates $\alpha_s^{(5)}$ to $\alpha_s^{\rm (SQCD)}$ & 2 \\
    \hline
  \end{tabular}
  \caption{\label{tab::CTs}Counterterms and decoupling constants
    entering the three-loop calculation of
  $C_1$.}
  \end{center}
\end{table}

For convenience we list in Tab.~\ref{tab::CTs} all counterterms and
decoupling constants needed
for our three-loop calculation together with the required loop order.

Let us remark that at LO and NLO the contributions to $C_1$ from diagrams as
shown in Fig.~\ref{fig::diags1} and in Fig.~\ref{fig::diags2} are separately
finite after taking into account all relevant counterterms. At NNLO, however,
the contribution from the top sector contains still $1/\epsilon$ poles
proportional to $M_Z^2/\msusy^2$ (where $\msusy$ denotes a squark mass) which
only cancel in combination with the result where the Higgs boson couples to a
super partner of one of the five light quarks (cf. Fig.~\ref{fig::diags2}).
Sample diagrams responsible for such poles are given in Fig.~\ref{fig::diags1}
(first diagram in second row) and Fig.~\ref{fig::diags2} (third diagram in
second row).

At this point one has arrived at a finite result for $C_1$ expressed in
terms of $\alpha_s^{(5)}$. We write its perturbative expansion 
in the form
\begin{eqnarray}
  C_1 &=&  - \frac{\alpha_s^{(5)}}{3\pi}\left( c_1^{(0)}(\mu_h)
    + \frac{\alpha_s^{(5)}(\mu_h)}{\pi} c_1^{(1)}(\mu_h)
    + \left(\frac{\alpha_s^{(5)}(\mu_h)}{\pi}\right)^2 c_1^{(2)}(\mu_h)
    + \ldots
  \right)\,,
\end{eqnarray}
where we have introduced the renormalization scale
$\mu_h$ (``h'' stands for ``hard'').
The one-loop expression is collected in $c_1^{(0)}(\mu_h)$.
In the SM and the MSSM it is given by
\begin{eqnarray}
  c_1^{(0),\rm SM} &=& 1\,,
\end{eqnarray}
and
\begin{eqnarray}
  c_1^{(0),\rm SQCD} &=& 
  \frac{1}{32}\CalphadivSbeta
  \Bigg[34 - \frac{\mstopone^2}{\mstoptwo^2} - \frac{\mstoptwo^2}{\mstopone^2}
  + \Cfourthetat\left(-2 + \frac{\mstopone^2}{\mstoptwo^2} +
    \frac{\mstoptwo^2}{\mstopone^2} \right) 
  + 8 \left(\frac{m_t^2}{\mstopone^2} + \frac{m_t^2}{\mstoptwo^2}\right)
  \Bigg]
  \nonumber\\&&\mbox{}
  + 
  \frac{M_Z^2\Salphapbeta}{16}\Bigg[
  \frac{2}{\msquark^2} - \frac{1}{\mstopone^2} - \frac{1}{\mstoptwo^2} 
  + \Ctwothetat\IEW\left(-\frac{1}{\mstopone^2} + \frac{1}{\mstoptwo^2}\right)
  \Bigg]
  \nonumber\\&&\mbox{}
  + 
  \frac{1}{4}
  \CalphadivSbeta\CSthetat\TalphapdivTbeta
  m_t\muSUSY
  \left(\frac{1}{\mstopone^2} - \frac{1}{\mstoptwo^2} \right)
  \,,
  \label{eq::C1SQCD}
\end{eqnarray}
respectively, where $c_x=\cos(x)$, $s_x=\sin(x)$, $t_x=\tan(x)$, $s_W$ is the
sine of the weak mixing angle $\theta_W$, $\muSUSY$ is the Higgs-Higgsino
bilinear coupling from the super potential and $\tan\beta$ is the ratio of the
vacuum expectation values of the two Higgs doublets.  The term proportional to
$M_Z^2/\msquark^2$ arises from the one-loop diagrams involving squarks, all
other terms origin from the top quark and top squark loops.
Note that Eq.~(\ref{eq::C1SQCD}) has been obtained with the help of the
relation
\begin{eqnarray}
  A_t &=& \frac{\mstopone^2-\mstoptwo^2}{2m_t}\sin(2\theta_t) + \muSUSY\cot\beta
  \,,
\end{eqnarray}
where $A_t$ is the soft SUSY breaking trilinear coupling of the top
squarks.

In the numerical results which are presented in Section~\ref{sec::num} we
include also effects from a non-vanishing bottom quark mass. The corresponding
formulae are taken over from the
literature~\cite{Degrassi:2010eu,Harlander:2010wr} and are included in our
program which computes the cross section. Thus, up to NLO we can study the
production of the light MSSM Higgs boson also for relatively small values of
$M_A$ and large $\tan\beta$, the region where the bottom quark contributions
are numerically important. At NNLO only those corrections are included which
survive in the limit $m_b\to 0$.


\subsection{\label{sub::appr}Constructing approximation results from expansions}

We evaluate the three-loop corrections in the following hierarchies
\begin{eqnarray}
  ({\rm h1}) && \msquark \approx \mstopone \approx \mstoptwo \approx \mgluino
  \gg m_t 
  \,,\nonumber\\
  ({\rm h2}) && \msquark \approx \mstoptwo \approx \mgluino \gg \mstopone \gg
  m_t 
  \,,\nonumber\\
  ({\rm h3}) && \msquark \approx \mstoptwo \approx \mgluino \gg \mstopone
  \approx m_t 
  \,,
  \label{eq::hier}
\end{eqnarray}
where $m_t$, $m_{\tilde{t}_{1,2}}$ and $m_{\tilde{g}}$ represent the top
quark, top squark, and gluino mass, respectively.  $m_{\tilde{q}}$ is
the generic mass of the supersymmetric partners of one of the five light
quarks which we assume degenerate. Thus we only allow for a mixing in the
top squark sector. A strong hierarchy between masses in
Eq.~(\ref{eq::hier}) suggests that we apply an asymptotic expansion (see
Ref.~\cite{Smirnov:2002pj} for a review) whereas a naive Taylor expansion in
mass differences is sufficient in case an approximation sign is present.
At three-loop order terms up to ${\cal O}(1/\msusy^{6})$ have been computed
for (h1) and (h3)\footnote{$\msusy$ is the common heavy mass in these
  hierarchies.} and up to ${\cal O}(1/\mstopone^{4})$ and ${\cal
  O}(1/\msusy^{4})$ for (h2). For each mass difference at least three expansion
terms (i.e. terms including $(m_i^2-m_j^2)^2$) have been evaluated.

In this context let us mention that in the hierarchies of Eq.~(\ref{eq::hier})
we have omitted the $\varepsilon$ scalar mass which is simultaneously
integrated out in order to guarantee the immediate matching to the Lagrangian
in Eq.~(\ref{eq::leff}) regularized using DREG.\footnote{In this context we
  also refer to the works~\cite{Bednyakov:2007vm,Bauer:2008bj,Kurz:2012ff}.}
We assume $M_\varepsilon$ to be much smaller than all other heavy masses which
facilitates the calculation. It is furthermore consistent with the condition
that the on-shell renormalized $\varepsilon$ scalar mass vanishes (see above).

Experience shows that the Taylor expansion around equal masses works quite
well and a few terms are sufficient in order to have good approximations for
$|m_1/m_2-1|<0.5$ (see, e.g., Refs.~\cite{Eiras:2006xm,Kant:2010tf} for
applications within a similar framework). This is the main motivation for
hierarchy (h1) which covers a big part of all interesting cases. We have
computed $C_1$ for the hierarchy (h2) in order to be able to cover the case
where one of the top squarks are light whereas the remaining supersymmetric
masses are above 1~TeV and close together. The motivation for (h3) comes from
situations of (h2) where $\mstopone$ and $m_t$ have the same order of
magnitude which is taken care of by a Taylor expansion in the mass difference.

The strong hierarchies in Eqs.~(\ref{eq::hier}) do not leave any freedom 
in view of the expansion parameter which is essentially given by the inverse
heavy mass. The choice of the mass in the numerator is practically irrelevant.
This is different for the expansions in mass differences. There is a vast
variety of choices which can lead to significant differences in the final
result. In the following we classify the various possibilities and construct a
procedure to obtain an optimized prediction.

The results for the hierarchies of Eq.~(\ref{eq::hier}) contain
either linear or quadratic differences of the involved masses, 
$m_{\tilde{t}_1}$, $m_{\tilde{t}_2}, m_{\tilde{q}}$ and $m_{\tilde{g}}$,
which we define as
\begin{eqnarray}
  \Delta_{ij}   &=& m_i-m_j\,,\nonumber\\
  \Delta^Q_{ij} &=& m_i^2-m_j^2\,,
\end{eqnarray}
respectively, and we consider expansions in
$\Delta_{ij}/m_i$ and $\Delta^Q_{ij}/m_i^2$.
For the evaluation of the integrals in the individual hierarchies 
we have chosen the following differences:
\begin{eqnarray}
  (\mbox{h1}): && 
  \Delta^Q_{\tilde{t}_1\tilde{t}_2}, 
  \Delta^Q_{\tilde{t}_1\tilde{q}},
  \Delta_{\tilde{t}_1\tilde{g}}\,,
  \nonumber\\
  (\mbox{h2}): &&
  \Delta^Q_{\tilde{g}\tilde{t}_2}, 
  \Delta^Q_{\tilde{g}\tilde{q}}\,,
  \nonumber\\
  (\mbox{h3}): && 
  \Delta^Q_{\tilde{t}_1 t},
  \Delta^Q_{\tilde{g}\tilde{t}_2}, 
  \Delta^Q_{\tilde{g}\tilde{q}}\,.
\end{eqnarray}
Note, however, this particular choice of representation may not be well suited
for the numerical evaluation of certain mass spectra.  For example, if
in the case of $(\mbox{h1})$ $m_{\tilde{t}_1}$ is the lightest supersymmetric
mass one observes a worse convergence as compared to a representation where
the expansion is carried out around one of the heavier masses, because higher
order terms in the original expansion are less suppressed.

The choice of the expansion point for the mass differences
is akin to the choice 
of the renormalization scale $\mu_R$. Often it is argued
that good choices go along with small higher-order corrections since certain
classes of corrections are automatically summed.  One could imagine a similar
mechanism when expanding in mass differences: a clever choice of the basis
mass can lead to small higher order terms in the expansion parameter. This
criterion will be used below for finding the best suited representation.

We have worked out two different approaches
to optimize the proper choice of the expansion parameters in an automatic way.

\begin{figure}[t]
  \centering
  \begin{tabular}{cccc}
    \includegraphics[width=0.2\linewidth]{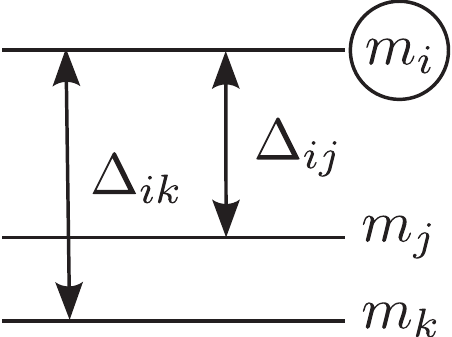} &
    \includegraphics[width=0.2\linewidth]{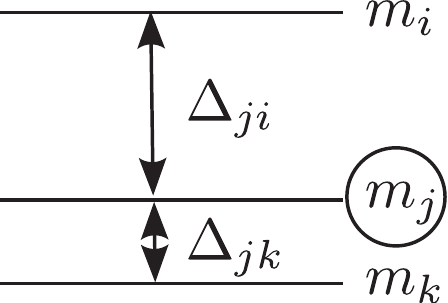} &
    \raisebox{-0.2cm}{\includegraphics[width=0.2\linewidth]
      {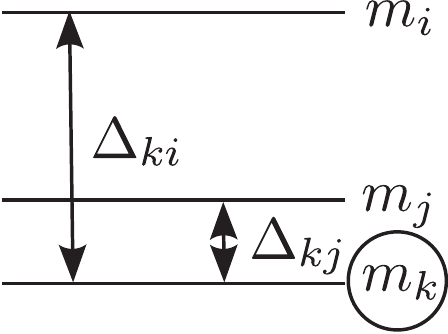}}
    &
    \raisebox{0.1cm}{\includegraphics[width=0.2\linewidth]
      {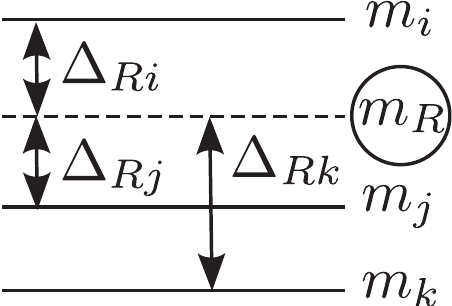}}\\
    (a) & (b) & (c) & (d) \\
  \end{tabular}
  \caption[]{\label{fig::massdiffsepctra}Figures (a) to (c) show different
    ways of expanding in small mass differences, when the masses $m_i, m_j$
    and $m_k$ are of the same order of magnitude.  In (a) the expansion
    is performed around the basis mass $m_i$, in (b) around $m_j$ and in (c)
    around $m_k$.  Figure (d) demonstrates the expansion around a generic
    reference mass $m_R$.  For the special values of $m_R$ equal to $m_i,
    m_j,$ or $m_k$ the representations of (a), (b) and (c) are reproduced.}
\end{figure}

In the first approach all possible representations of the underlying expansion
are generated automatically both for linear or quadratic mass differences
$\Delta_{ij}$ and $\Delta^Q_{ij}$.  To this end in a first step the existing
representation is rewritten in terms of new mass differences for each
possible choice of the basis mass which is illustrated in
Fig.~\ref{fig::massdiffsepctra}(a)--(c).  Afterwards an expansion in small
quantities is performed respecting the original expansion depths.  This
approach has already been introduced in Ref.~\cite{Zerf:2012}.

For the hierarchy $(\mbox{h1})$ we have four different basis masses
($m_{\tilde{g}}, m_{\tilde{t}_1}, m_{\tilde{t}_2}, m_{\tilde{q}}$) and 
two possibilities (linear, quadratic) to write each of the three mass
differences. Thus one has $4\times 2^{3}=32$ different representations.  For
the hierarchy $(\mbox{h2})$ we have $3\times 2^{2}=12$ different
representations and for $(\mbox{h3})$ one obtains $(2\times
2^{1})\times(3\times 2^{2})=48$ different representations where the first
bracket results from the expansion $m_t\approx m_{t_1}$ and the second one
from $m_{\tilde{t}_2} \approx m_{\tilde{q}} \approx m_{\tilde{g}}$.

After generating all representations for a certain hierarchy they are
evaluated numerically for a given supersymmetric spectrum.  In order to select
the representation which leads to the best approximation we define two
quantities which we require to be small (see also
Ref.~\cite{Kurz:2012ff}). The first one controls the deviation to the exact
two-loop result and is defined through
\begin{equation}
  \delta_1 = 
  \frac{c_1^{(1), \text{app}}-c_1^{(1), \text{exact}}}{c_1^{(1),
      \text{exact}}}
  \,,
  \label{eq::delta1}
\end{equation}
where $c_1^{(1), \text{app}}$ contains as many expansion terms as the
three-loop result we are interested in.
The second quantity is given by
\begin{equation}
  \delta_2 = 
  \frac{c_1^{(2), \text{app,c}}-c_1^{(2), \text{app}}}{c_1^{(2), \text{app}}}
  \,,
\end{equation}
where the superscript ``c'' indicates that the expansion is mass
  differences is reduced from three to two terms.
$\delta_2$ is a measure for the importance of
higher order terms and thus the convergence of the expansions.

The best representation is chosen by minimizing 
the combination
\begin{equation}
  \delta_{\rm app} = \sqrt{ |\delta_1|^2 + |\delta_2|^2 } \,,
\end{equation}
i.e., we require both good agreement with the exact result at two loops
and a good convergence within the three-loop expression.

Our second method is a generalization of the first one in the sense that for
the basis mass we do not use one of the physical masses involved in the
calculation but a generic reference mass $m_R$.  Of course, the numerical
value of $m_R$ should be of the order of the other masses, however, apart from
that there is no restriction.  This situation is illustrated in
Fig.~\ref{fig::massdiffsepctra}(d).
Choosing a variable (although definite) basis mass reduces the number of
representations, $N_R$, which can be seen from the following formulae
\begin{eqnarray}
  N^{(\text{fbm})}_R &=& \prod_{i=1}^{i_{\text{max}}}n_i2^{n_i-1}\,,\nonumber\\
  N^{(\text{vbm})}_R &=& \prod_{i=1}^{i_{\text{max}}}2^{n_i}\,.
  \label{eq::NR}
\end{eqnarray}
where ``fbm'' (fixed-basis-mass) and ``vbm'' (variable-basis-mass) refers to
the first and second approach, respectively. In Eq.~(\ref{eq::NR})
$i_{\text{max}}$ refers to the number of independent regions with expansions
in mass differences (for $(\mbox{h1})$ and $(\mbox{h2})$ we have
$i_{\text{max}}=1$ and for $(\mbox{h3})$ $i_{\text{max}}=2$) and $n_i$ to the
number of physical particle masses involved in the $i$-th region.  In the
``vbm'' scheme this leads to 16, 8 and 32 different representations for the
hierarchies (h1), (h2) and (h3), respectively.

Let us next discuss the criteria which fix $m_R$ in the ``vbm'' scheme.  
Note that the exact two-loop result is independent of $m_R$ and
thus $m_R$ can be chosen in such a way that the approximated two-loop
expression has minimal deviation from the exact result.  This means that one
has to find the minimum of the function $|\delta_1(m_R)|$ (see
Eq.~(\ref{eq::delta1})) which leads to the reference mass
$m_{R}^{(\delta_1)}$. The latter can be used to obtain the three-loop term of
the matching coefficient, $c_1^{(2)}(m_{R}^{(\delta_1)})$. Altogether we have
\begin{equation}
  |\delta_1(m_{R})| \overset{!}{=} \text{min}
  \quad\Rightarrow \quad {m_{R}^{(\delta_1)}}
  \quad\Rightarrow \quad c_1^{(2)}(m_{R}^{(\delta_1)})
  \,,
  \label{eq::del1mR}
\end{equation}
where for the starting value of $m_R$ we choose the
average over all involved masses.

In complete analogy we determine the quantities 
$m_{R}^{(\delta_2)}$, $m_{R}^{(\delta_{1+2})}$,
$c_1^{(2)}(m_{R}^{(\delta_2)})$ and $c_1^{(2)}(m_{R}^{(\delta_{1+2})})$:
\begin{align}
  |\delta_2(m_{R})| &&\overset{!}{=}&& \text{min}
  &&\quad\Rightarrow&&& {m_{R}^{(\delta_2)}}
  &&\quad\Rightarrow&&& c_1^{(2)}(m_{R}^{(\delta_2)})
  \,,\nonumber\\
  |\delta_1(m_{R})|+ |\delta_2(m_{R})| &&\overset{!}{=}&& 
  \text{min}
  &&\quad\Rightarrow&&& {m_{R}^{(\delta_{1+2})}}
  &&\quad\Rightarrow&&& c_1^{(2)}(m_{R}^{(\delta_{1+2})})
  \,.
  \label{eq::del12mR}
\end{align}
For well-converging expansions we expect similar values of $m_{R}^{(\delta)}$
for all three choices of $\delta$.

The information collected so far is used to compute the uncertainty of the 
approximation for $c_1^{(2)}$. We define it to be proportional to the 
deviations of the approximated results based on the criteria
in Eqs.~(\ref{eq::del1mR}) and~(\ref{eq::del12mR}) to their
average. Since this term could by chance be small an additional factor is
introduced which takes into account the size of $\delta_1$ and $\delta_2$.
Our final formula for the uncertainty reads
\begin{equation}
  \Delta c_1^{(2)} = 
  \left[1+\left|\delta_1(m_{R}^{(\delta_{1+2})})\right|
  +\left|\delta_2(m_{R}^{(\delta_{1+2})})\right|\right]
  \sum_{\delta=\delta_1,\delta_2,\delta_{1+2}}
  \left|c_1^{(2)}(m_{R}^{(\delta)})-\overline{c}_1^{(2)}\right|
  \,,
  \label{eq::Delta_c}
\end{equation}
where $\overline{c}_1^{(2)}$ is given by:
\begin{equation}
  \overline{c}_1^{(2)}=\frac{1}{3}\sum_{\delta=\delta_1,\delta_2,\delta_{1+2}}
  c_1^{(2)}(m_{R}^{(\delta)})\,.
\end{equation}

The described procedure leads to predictions for the three-loop coefficient in
the form $c_1^{(2)}(i)\pm \Delta c_1^{(2)}(i)$ where $i$ runs over all
representations. For our final result we pick the one with the smallest error,
after ensuring that all predictions agree within their errors. In order to
fulfill the latter step it may happen that the uncertainties have to be
rescaled.  Note that in all results presented in Section~\ref{sec::num}
$\Delta c_1^{(2)}$ is negligibly small.


\subsection{\label{sub::c12l}$C_1$ to two loops}

In this subsection we discuss the application of the formalism described above
to the two-loop result for $C_1$ where the occurring momentum integrals have
been evaluated both by taking into account the exact dependence on all mass
parameters and by applying the hierarchies defined in Eq.~(\ref{eq::hier}).  A
detailed numerical discussion of the cross section will be presented in
Section~\ref{sec::num}. In the following we want to test the approximation
procedure and discuss its accuracy.  For this purpose we adopt the following
values for the input parameters
\begin{align}
  &&m_t^{\overline{\rm DR}}=155~\mbox{GeV}\,,
  &&A_t=100~\mbox{GeV}\,,
  &&\tilde{m}_{Q_3}=500~\mbox{GeV}\,,
  \nonumber\\
  &&\muSUSY=100~\mbox{GeV}\,,
  &&\tan\beta = 10\,.
  \label{eq::parameters}
\end{align}
where $A_t$ is the trilinear coupling, $M_Z$ is the $Z$ boson mass and
$\tilde{m}_{Q_3}$, a soft SUSY breaking parameter for the squark doublet of the
third family.  We furthermore fix the renormalization scale to
$\mu_R=M_t$, where $M_t$ is the on-shell top quark mass.

\begin{figure}[t]
  \centering
  \includegraphics[width=.8\linewidth]{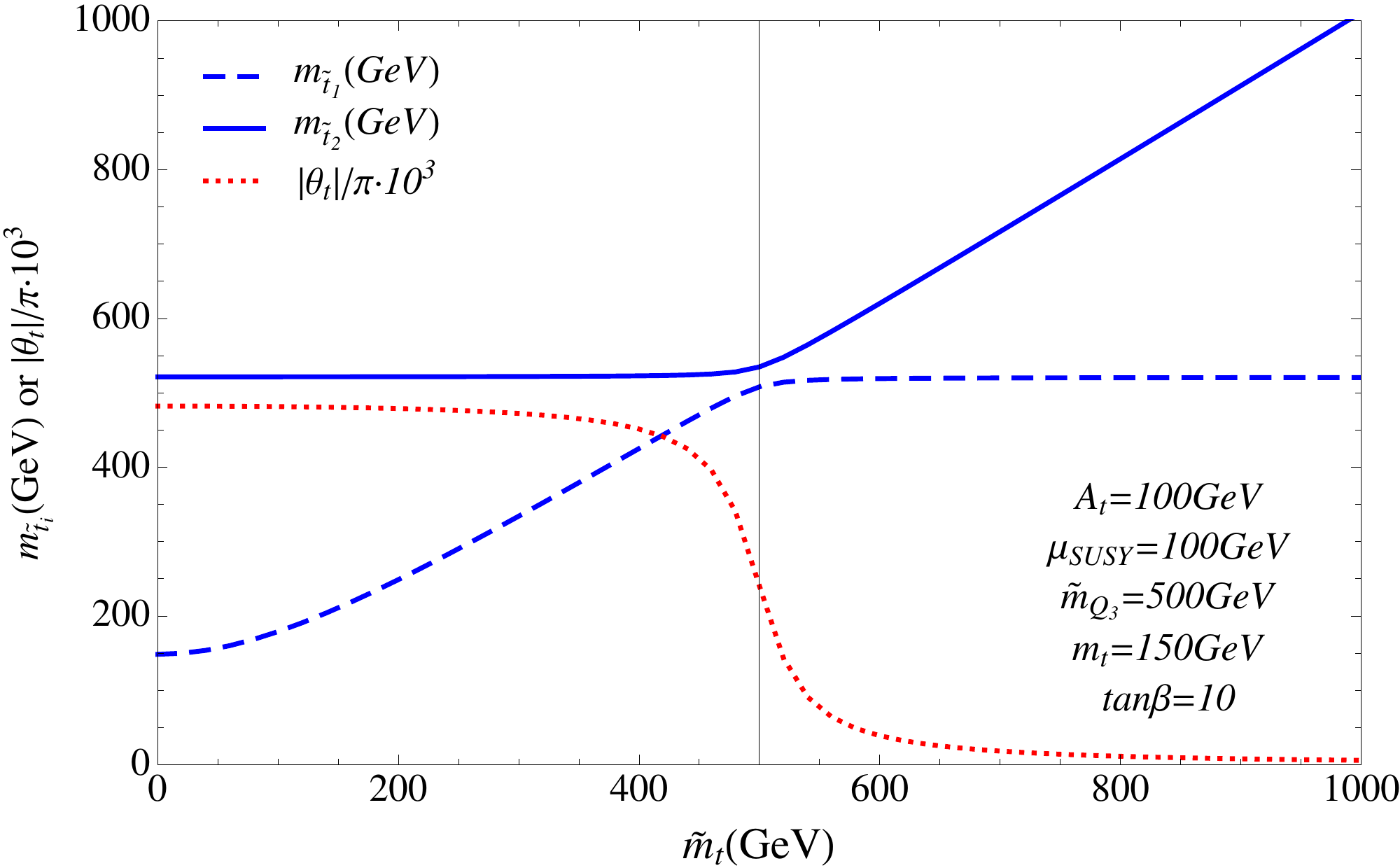}
  \caption[]{\label{fig::mst12thet} $\mstopone$, $\mstoptwo$ and $\theta_{t}$
    obtained from the diagonalization of the top squark mass matrix as a
    function of the soft SUSY breaking parameter $\mtildeuR$.}
\end{figure}

Once these parameters are fixed it is possible to compute the physical masses
$\mstopone$, $\mstoptwo$ and $\theta_{t}$ in dependence on the parameter
$\mtildeuR$, the singlet soft SUSY breaking mass parameter of the right-handed
top squark.\footnote{See Ref.~\cite{Martin:1997ns} for details on the
  notation.}  This is achieved by diagonalizing the corresponding mass matrix.
For convenience we show in Fig.~\ref{fig::mst12thet} $\mstopone$, $\mstoptwo$
and $\theta_{t}$ for $\mtildeuR$ between 0 and 1~TeV.  For small and large
values of $\mtildeuR$ one obtains a large splitting in the top squark masses
whereas for $\mtildeuR\approx 500$~GeV they are approximately degenerate.

\begin{figure}[t]
  \centering
  \begin{tabular}{c}
    \includegraphics[width=.8\linewidth]{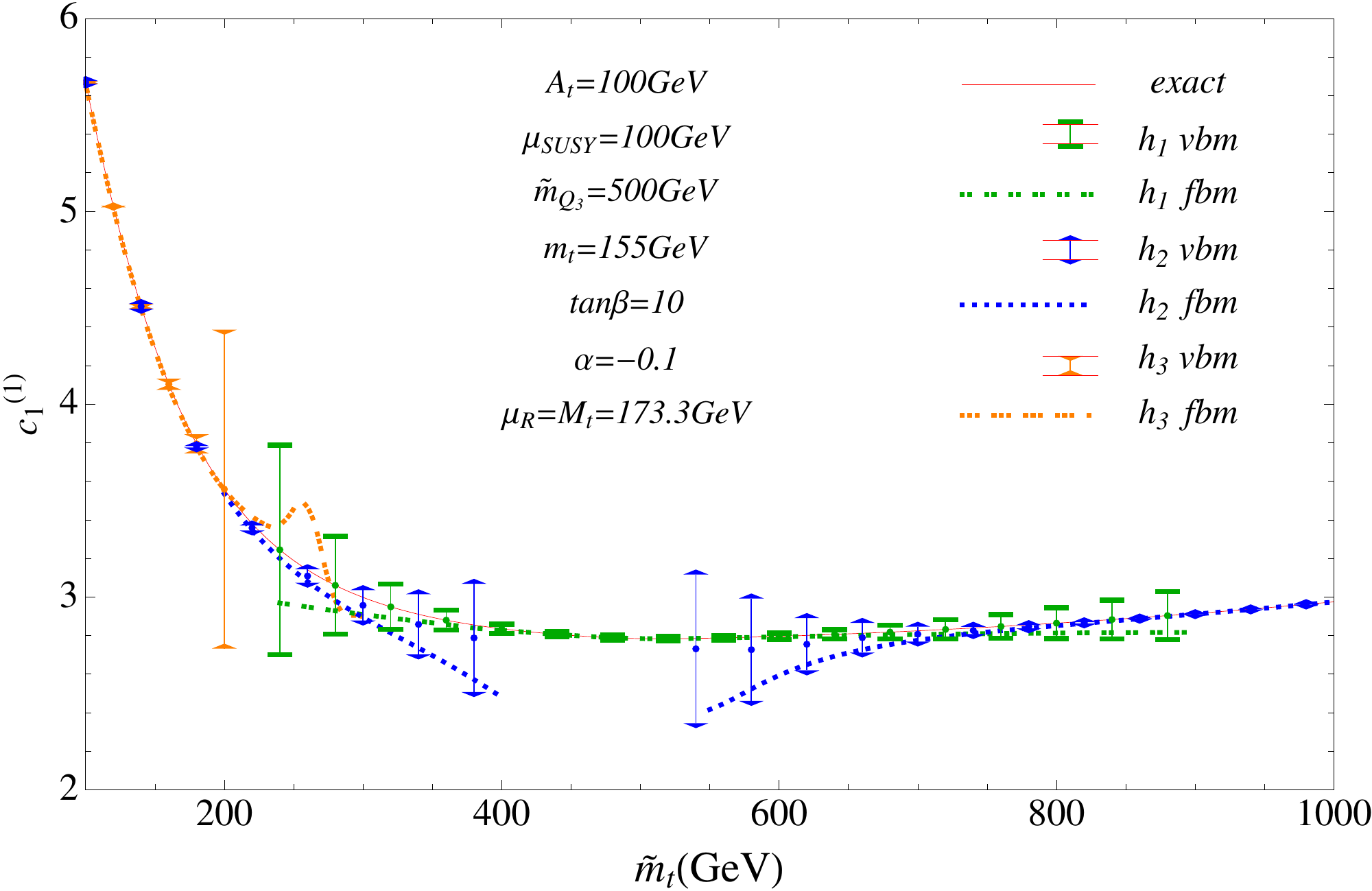}
  \end{tabular}
  \caption[]{\label{fig::c12l}Two-loop correction to $C_1$ as a function of 
    $\mtildeuR$. See text for further details.}
\end{figure}

The numerical result for $c_1^{(1)}$ is shown in Fig.~\ref{fig::c12l}
which is
obtained by using $\mstoptwo=\msquark=\mgluino$ and $\alpha=-0.1$.
Note that the hierarchies
(h3) and (h2) are designed to lead to good approximations for small and large
$\mtildeuR$ whereas (h1) should lead to good results for intermediate
values.  In Fig.~\ref{fig::c12l} the exact result is shown as solid line and
the approximations for the hierarchies (h1), (h2) and (h3) as dashed
lines, where we also plot the
corresponding uncertainties obtained from the ``vbm'' method described in
Subsection~\ref{sub::appr}. 
For clarity, the curves are truncated beyond the points where
the corresponding uncertainty grows too large.
The dotted lines correspond to
the ``fbm'' method. They are
different from the dashed ones only for a few values of $\mtildeuR$.
From Fig.~\ref{fig::c12l} one observes that
for each value of $\mtildeuR$ there is at least one hierarchy which leads
to an excellent approximation of the exact curve. In particular, good
agreement is observed in the overlapping regions around
$\mtildeuR\approx 300$~GeV and 600~GeV.  Furthermore, the hierarchy (h2)
leads to good results also for small values of $\mtildeuR$ which is
not expected in the first place. It seems, however, that the expansion for
$\mstopone\gg m_t$ converges well. Moreover, one can learn from
Fig.~\ref{fig::c12l} that the approximation procedure seems to work very well
since the estimated uncertainties cover the differences between the different
hierarchies.

Let us briefly comment on the limit $\mtildeuR\to 0$ where 
$c_1^{(1)}$ becomes numerically large.
It originates from contributions containing self
energy corrections to the $\tilde{t}_1$ where in the loop heavy supersymmetric
masses are present. In the $\overline{\rm DR}$ scheme this leads to
enhancement factors proportional to $\msusy^2/\mstopone^2$ which become
large for small values of $\mstopone$. As a way out it is possible to use the
$\overline{\rm MDR}$ scheme introduced in Ref.~\cite{Kant:2010tf} which
systematically subtracts these terms. In the phenomenological results
presented in Section~\ref{sec::num} the supersymmetric parameters are such that 
the factors $\msusy^2/\mstopone^2$ are not numerically dominant and thus
we are not forced to introduce the $\overline{\rm MDR}$ scheme.

At this point a detailed comparison to {\tt evalcsusy} is in
order.\footnote{Note that the authors of Ref.~\cite{Degrassi:2008zj} found
  agreement with the results of Ref.~\cite{Harlander:2004tp}.} In
Ref.~\cite{Harlander:2004tp} DRED has been implemented in a naive way by
performing the Lorentz algebra in four and the loop integrals in
$D=4-2\epsilon$ dimensions, no $\varepsilon$ scalars have been introduced.
Furthermore, {\tt evalcsusy} only contains contributions where the Higgs boson
couples to a top quark or top squark, diagrams as the ones shown in
Fig.~\ref{fig::diags2} are not included. Thus the result for $C_1$ contains
terms of order $1$, $m_t^2/m_{\tilde{t}}^2$, $\mu_{\rm susy}
m_t/m_{\tilde{t}}^2$ and $M_Z^2/m_{\tilde{t}}^2$ (as can also be deduced from
the structure of the couplings as displayed in the Appendix of
Ref.~\cite{Harlander:2004tp}). As already mentioned above, the
$M_Z^2/m_{\tilde{t}}^2$ term is numerically small.
The NLO calculation performed in this paper reproduces all
terms contained in {\tt evalcsusy}.

Let us stress that it is not possible to apply the naive 
approach of Ref.~\cite{Harlander:2004tp} at NNLO.


\subsection{$C_1$ to three loops}

\begin{figure}[t]
  \centering
  \begin{tabular}{c}
    \includegraphics[width=.8\linewidth]{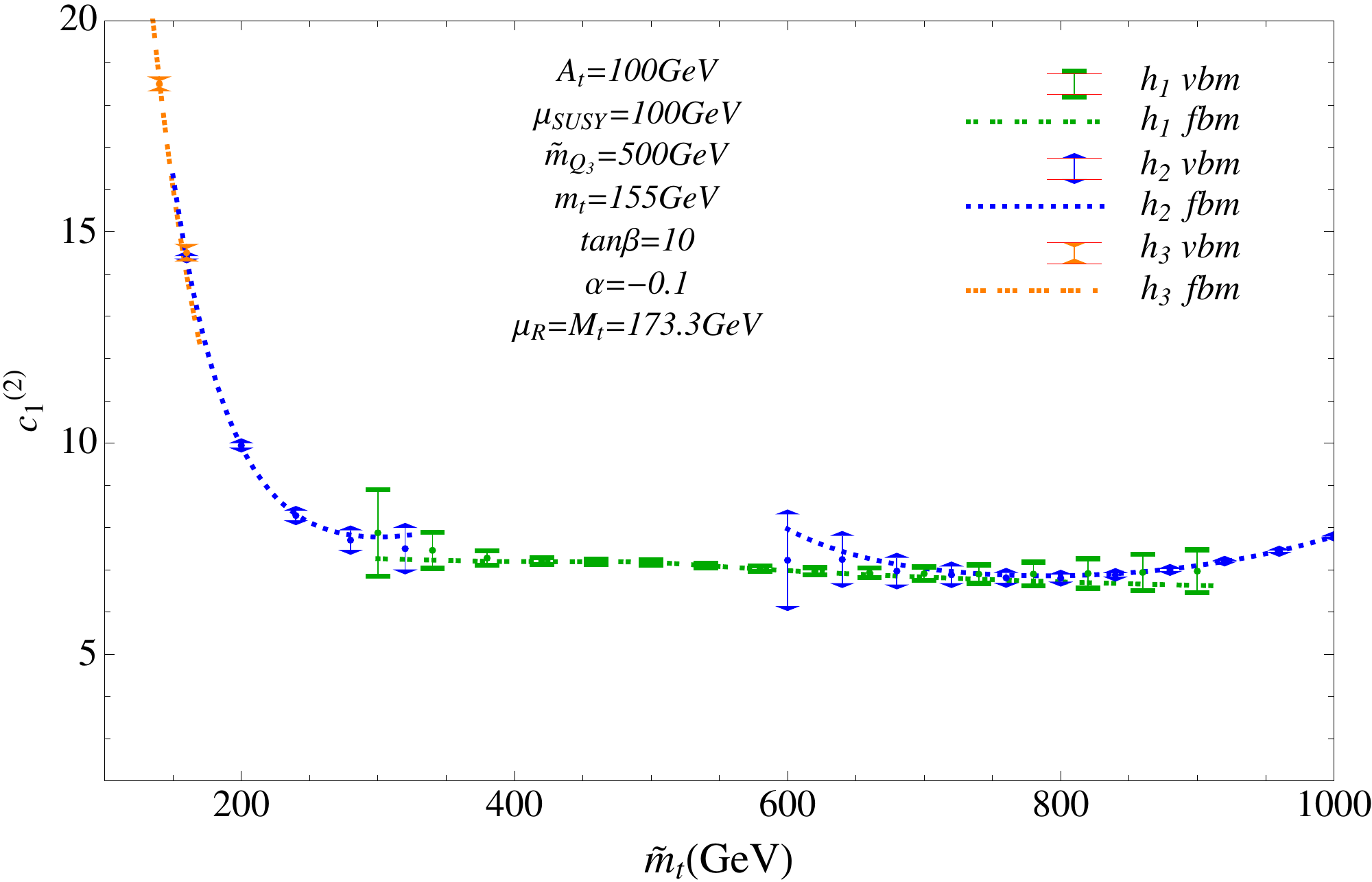}
  \end{tabular}
  \caption[]{\label{fig::c13l}Three-loop correction to $C_1$ as a function of 
    $\mtildeuR$. See text for further details.}
\end{figure}

In this subsection we use the parameters of Eq.~(\ref{eq::parameters})
and the procedure described below this equation to evaluate the three-loop
corrections to $C_1$. The result is shown in Fig.~\ref{fig::c13l} where the
same notation as at two loops (see Fig.~\ref{fig::c12l}) has been used.
The qualitative behaviour of the three-loop curves is very similar to the
corresponding two-loop results in Fig.~\ref{fig::c12l}. Again one observes
a nice overlap between the different hierarchies such that the whole
parameter region can be covered.

The analytical expressions are again quite lengthy and thus we refrain
from presenting all of them in the paper but provide computer-readable
expressions~\cite{progdata}. To illustrate the structure of the results
we list $C_1$ for the hierarchy (h1) in the degenerate mass limit. It
reads
\begin{eqnarray}
  C_1^{\overline{\rm DR},(\rm h1)} &\!\!\!\!=\!\!\!&
  -\frac{\alpha_s^{(5)}}{3\pi}\frac{c_\alpha}{s_\beta}
  \Bigg\{
  1 + \frac{m_t^2}{2\msusy^2} 
  + \frac{\alpha_s^{(5)}}{\pi}\Bigg[
  \frac{11}{4} - \frac{1}{3}\left(t_\alpha +
    \frac{1}{t_\beta}\right)\xmus 
  + \left(\frac{23}{12} + \frac{5}{12}\lmMsusy 
    \right.\nonumber\\&&\left.\mbox{}
    -\frac{5}{12}\lmMt\right)\frac{m_t^2}{\msusy^2}
  - \iCalphadivSbeta\Salphapbeta\frac{M_Z^2}{9\msusy^2}
  \Bigg]
  \!+ \!\left(\frac{\alpha_s^{(5)}}{\pi}\right)^2
  \!\Bigg[
  \frac{2777}{288} + \frac{19}{16}\lmMt + n_l \left(-\frac{67}{96} +
    \frac{1}{3}\lmMt\right)
  \nonumber\\&&\mbox{}
  + \left(-\frac{85}{54} - \frac{85}{108}\lmMsusy -
    \frac{1}{18}\lmMt  + n_l\left(-\frac{1}{18} +
      \frac{1}{12}\lmMsusy\right)
  \right)\left(t_\alpha+\frac{1}{t_\beta}\right)\xmus
  \nonumber\\&&\mbox{}
  + \Bigg(
  \frac{30779857}{648000} - \frac{6029}{10800}\lmMsusy 
  - \frac{475}{288}\lmMsusy^2 - \frac{20407}{10800}\lmMt 
  + \frac{26}{45}\lmMsusy\lmMt - \frac{377}{1440}\lmMt^2 
  \nonumber\\&&\mbox{}
  - \frac{15971}{576}\zeta_3 
  + n_l\left(
    - \frac{3910697}{216000} 
    + \frac{63}{4}\zeta_3
    + \frac{3307}{4800}\lmMsusy
    + \frac{131}{576}\lmMsusy^2
    + \frac{5113}{14400}\lmMt
    \right.\nonumber\\&&\left.\mbox{}
    - \frac{101}{1440}\lmMsusy\lmMt
    + \frac{3}{320}\lmMt^2
  \right)
  \Bigg)
  \frac{m_t^2}{\msusy^2}
  + \iCalphadivSbeta\Salphapbeta
  \frac{M_Z^2}{\msusy^2}
  \Bigg(-\frac{209}{432} + \frac{1}{12}\lmMsusy - \frac{1}{54}\lmMt
  \nonumber\\&&\mbox{}
    + n_l \left(-\frac{1}{72} + \frac{1}{216}\lmMsusy\Bigg)
  \right)
  \Bigg]
  + {\cal O}\left(\frac{m_t^3}{\msusy^3}\right)
  \Bigg\}
  \,.
  \label{eq::C1DR}
\end{eqnarray}
As compared to Ref.~\cite{Pak:2010cu} the $M_Z^2$ terms, which result both
from the top quark/top squark and light quark/squark sector, are new.



\section{\label{sec::num}Numerical predictions}


\subsection{\label{sub::form}Formalism}

As far as the contribution to the cross section originating from the top
quark, top squark and squark sector is concerned we adopt the formalism
outlined in Ref.~\cite{Pak:2010cu}.  In particular, we apply a redefinition of
the effective-theory operator in such a way that both the effective operator
and the redefined Wilson coefficient are renormalization group invariant. This
allows us to separate the renormalization scale into a hard part, which enters
the matching coefficient, and a soft part present in the matrix elements
computed with the effective operator.  In this way potentially large
logarithms $\ln(\mu_h/M_h)$ are resummed where $\mu_h$ is of the order of the
top quark or SUSY particle masses.  In this framework the cross section can be
computed from the following formula:
\begin{eqnarray}
  \sigma^{\rm SQCD}_t(\mu_s,\mu_h) &\!\!\!=\!\!\!& \sigma_0
  \Bigg\{\Sigma^{(0)} 
    + \frac{\alpha_s^{(5)}(\mu_s)}{\pi}\Sigma^{(1)} 
    + \frac{\alpha_s^{(5)}(\mu_h)}{\pi}2c_g^{(1)}\Sigma^{(0)} 
    + \left(\frac{\alpha_s^{(5)}(\mu_s)}{\pi}\right)^2\Sigma^{(2)} 
    \label{eq::sigma_t}
    \\&&\mbox{}
    + \frac{\alpha_s^{(5)}(\mu_s)}{\pi}\frac{\alpha_s^{(5)}(\mu_h)}{\pi}
    2c_g^{(1)}\Sigma^{(1)}  
    + \left(\frac{\alpha_s^{(5)}(\mu_h)}{\pi}\right)^2
    \left[(c_g^{(1)})^2+2c_g^{(2)}\right]\Sigma^{(0)}   
    + \ldots
    \Bigg\}
    \,,
    \nonumber
\end{eqnarray}
where 
\begin{eqnarray}
  C_g &=& 1 + \frac{\alpha_s^{(5)}(\mu_h)}{\pi} c_g^{(1)}
  + \left(\frac{\alpha_s^{(5)}(\mu_h)}{\pi}\right)^2
  c_g^{(2)}
  + \ldots
  \,,\nonumber\\
  \Sigma &=& \Sigma^{(0)} + \frac{\alpha_s^{(5)}(\mu_s)}{\pi} \Sigma^{(1)}
  + \left(\frac{\alpha_s^{(5)}(\mu_s)}{\pi}\right)^2 \Sigma^{(2)}
  + \ldots  \,,
\end{eqnarray}
and
\begin{align}
 \sigma_0=&\frac{9}{4}\Bigg|\frac{c_\alpha}{s_\beta}I(\tau_t)
               +\sum_{i=1,2}S_i\bigg\{
                    \frac{c_\alpha}{s_\beta}
                    c_{\theta_t}s_{\theta_t}\bigg(t_{\alpha}+\frac{1}{t_{\beta}}\bigg)\frac{m_t\mu_{\rm susy}}{2m_{\tilde{t}_i}^2}
                    +\frac{c_\alpha}{s_\beta}
                    \frac{m_t^2}{8 m_{\tilde{t}_i}^2}\bigg[s_{2\theta_t}^2\frac{m_{\tilde{t}_1}^2-m_{\tilde{t}_2}^2}{m_t^2}-4S_i\bigg]\nonumber\\&
                    +s_{\alpha+\beta}\frac{M_Z^2}{8 m_{\tilde{t}_i}^2}\bigg[c_{2 \theta_t}\bigg(\frac{8}{3}s_W^2-1\bigg)+S_i\bigg]
               \bigg\}\tilde{I}(\tau_{\tilde{t}_i})
               -s_{\alpha+\beta}\frac{M_Z^2}{4m_{\tilde{q}}^2}\tilde{I}(\tau_{\tilde{q}})\Bigg|^2\,,
\end{align}
with
\begin{align}
  \tau_i=&\frac{4 m_i^2}{M_h^2}\,,\quad  S_1=-1\,,\quad  S_2=+1\,, \nonumber\\
  I(\tau)=&\tau \big[1+\big(1 - \tau\big) f(\tau) \big]\,,\quad
  \tilde{I}(\tau)=\tau \big(1 - \tau f(\tau)\big)\,,\nonumber\\
  f(\tau) =& 
  \left\{\begin{array}{cl}
      \arcsin^2(1/\sqrt{\tau}), & \tau \ge 1\,, \\
      - \frac{1}{4} \left(\ln\frac{1 + \sqrt{1-\tau}}{1 - \sqrt{1-\tau}} - i\pi \right)^2,
      & \tau < 0\,. 
    \end{array}
  \right.
\end{align}
The coefficient function
$C_g$ is obtained from the renormalized Wilson coefficient $C_1$ in
Eq.~(\ref{eq::leff}) via
\begin{eqnarray}
  C_g &=& -\frac{3\pi}{ c_1^{(0)} } \frac{1}{B^{(5)}} C_1 \,,
  \label{eq::B5}
\end{eqnarray}
with
\begin{eqnarray}
  B^{(5)} &=& -\frac{\pi^2 \beta^{(5)}}{\beta_0^{(5)}\alpha_s^{(5)}}
  \,.
\end{eqnarray}
In Eq.~(\ref{eq::B5}) we divide by $c_1^{(0)}$ since it is already included in
$\sigma_0$ which is factored out from Eq.~(\ref{eq::sigma_t}).
Note that the coefficients $c_g^{(1)}$ and $c_g^{(2)}$ depend on $\mu_h$ which
is suppressed in Eq.~(\ref{eq::sigma_t}).

$\Sigma^{(k)}$ depends on $\mu_s$ and represents the N$^k$LO contribution of
the hadronic cross section computed within five-flavour QCD using the
effective Lagrange density in Eq.~(\ref{eq::leff}) where a summation over all
sub channels ($gg, qg, q\bar{q}, qq$ and $qq^\prime$) is understood.  Note
that this part can be taken over from the corresponding SM calculation; in our
case we use the results from Ref.~\cite{Pak:2011hs}.  In principle
$\Sigma^{(k)}$ also depends on the factorization scale $\mu_f$, however, in
our calculation we identify $\mu_f$ and $\mu_s$.  $\Sigma$ is by construction
renormalization group invariant since $B^{(5)}$ is incorporated.  Note that
for $\mu_h=\mu_s$ $B^{(5)}$ cancels in the above formulae.

The approach described so far is not practical in case the bottom quark mass
is kept different from zero since then a division into $\mu_h$ and $\mu_s$ is
not as obvious as in the top and squark sector where the calculation is based
on the effective theory.  Thus the numerical evaluation of the cross section
including finite bottom quark mass effects proceeds as follows: we set
$\mu_h=\mu_s$ and evaluate the cross section up to NLO using the expressions
from Ref.~\cite{Degrassi:2010eu}.  Afterwards we subtract our NLO top- and
squark-sector contribution based on Eq.~(\ref{eq::sigma_t}) choosing also
$\mu_h=\mu_s$ and finally
add our NNLO result as given in Eq.~(\ref{eq::sigma_t}).
At the end we include electroweak corrections assuming complete
factorization. Thus our best prediction looks as follows
\begin{eqnarray}
  \label{eq::sigma}
  \sigma^{\rm SQCD}(pp\to h+X) &=& \left( 1 + \delta^{\rm EW} \right) 
  \times
  \\&&
  \Bigg[\sigma_{tb}^{\rm SQCD}(\mu_s)\bigg|_{\rm NLO}
    -   \sigma_t^{\rm SQCD}(\mu_s)\bigg|_{\rm NLO}
    +   \sigma_t^{\rm SQCD}(\mu_s,\mu_h)\bigg|_{\rm NNLO}
  \Bigg]
  \nonumber
  \,,
\end{eqnarray}
where $\delta^{\rm EW}$ is taken from Ref.~\cite{Actis:2008ug}
(see also Ref.~\cite{Anastasiou:2008tj}).
Note that in
order to obtain the NNLO prediction the convolution of
all contributions entering
Eq.~(\ref{eq::sigma}) is performed with NNLO
parton distribution functions (PDFs).
This is also true for the quantities $\sigma_{tb}^{\rm
  SQCD}(\mu_s)|_{\rm NLO}$ and $\sigma_t^{\rm SQCD}(\mu_s)|_{\rm NLO}$.
Analogous formulae to Eq.~(\ref{eq::sigma}) are also
used for LO and NLO predictions of $\sigma(pp\to h+X)$. Of course, in these cases
LO and NLO PDFs are used, respectively.


\subsection{Numerical setup}

\begin{figure}[t]
  \centering
  \includegraphics[width=\linewidth]{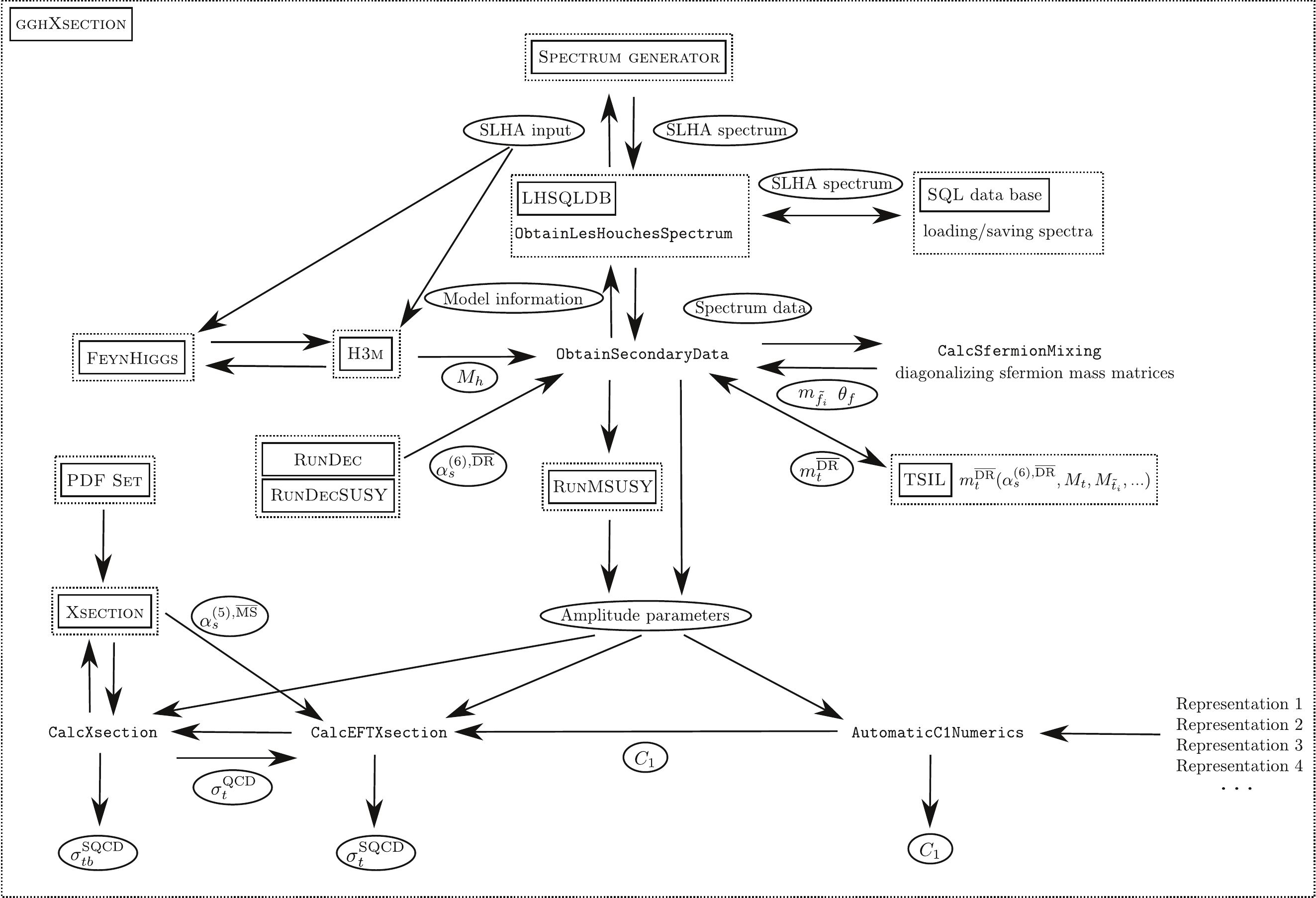}
  \caption[]{\label{fig::fchart}Flow chart of our numerical setup.
    The rectangular frames denote stand-alone program packages which are
    combined by {\tt gghXsection}. It controls the flow of data
    (indicated by the ovals) among the individual components.
    More details are given in the text.}
\end{figure}

In the following we briefly describe the important features of our numerical
program {\tt gghXsection}. 
It consists of several components which take care of the generation
of the spectrum, the computation of the Higgs boson mass,
the convolution of the partonic cross sections 
with the  PDFs
in the effective theory and the determination of the best approximation for the
three-loop matching coefficient. A flow chart showing the individual
components can be found in Fig.~\ref{fig::fchart}.
\begin{itemize}

\item The individual components are called and controlled by the {\tt
    Mathematica} program {\tt gghXsection}.

\item Our analytical expressions are parametrized in terms of $\overline{\rm
    DR}$ mass parameters defined at the renormalization scale $\mu_R$.  For an
  automated generation of the required MSSM parameter spectra within {\tt
    Mathematica} we use our own {\tt Mathematica} package {\tt
    LHSQLDB}.\footnote{LHSQLDB stands for {\bf L}es {\bf H}ouches {\bf S}pectra
    \& {\bf SQL} {\bf D}ata {\bf B}ase, where the two S are melted to one.}
  This package works as interface between {\tt Mathematica} and the SUSY spectrum
  generators following the ``Les Houches Accord''
  \cite{Skands:2003cj,Allanach:2008qq} like {\tt
    \softsusy}~\cite{Allanach:2001kg}, {\tt Spheno}~\cite{Porod:2003um} or {\tt
    Suspect}~\cite{Djouadi:2002ze} (we use {\tt \softsusy} as default spectrum
  generator throughout this paper).  Furthermore it is able to connect {\tt
    Mathematica} to an SQL data base where SUSY spectra can be automatically
  saved to and/or loaded from, avoiding multiple recalculations of already
  known parameters.
  
  Since this package might be of general use, a separate publication is
  planned.  However, a private version can be obtained from the authors upon
  request.

\item
  We define the strong coupling constant in
  five-flavour QCD using the $\overline{\rm MS}$ scheme. The numerical 
  results for $\alpha_s^{(5)}(\mu_R)$ are taken from the 
  PDF sets.

\item The spectrum generator also provides input for {\tt
    H3m}~\cite{Harlander:2008ju,Kant:2010tf} which is used to compute the
  light MSSM Higgs boson mass to three-loop accuracy.  {\tt H3m} uses
    {\tt FeynHiggs}~\cite{Frank:2006yh} for the two-loop expressions.

  Note that the spectrum generators can in general only be used for
  renormalization scales as low as the $Z$ boson mass. However, in our
  numerical analyses also lower scales appear since we choose $\mu_s=M_h/2$ as
  a central value. For this reason we have implemented the renormalization
  group equations (taken from Ref.~\cite{Avdeev:1997vx}) ourselves in order to
  obtain the spectrum for all desired values of $\mu_s$  (see the program
  {\tt RunMSUSY} in Fig.~\ref{fig::fchart}).

\item In order to compute the $\overline{\rm DR}$ top quark mass from the
  on-shell one we use the results of Ref.~\cite{Martin:2005ch,Martin:2005qm}
  which are based on the program {\tt TSIL}.  
  For this step
  we need the strong coupling constant in the full theory which we obtain with
  the help of {\tt RunDec}~\cite{Chetyrkin:2000yt,Schmidt:2012pw} and the
  extension of the decoupling and running routines to supersymmetry based on
  Refs.~\cite{Harlander:2005wm,Bauer:2008bj}.

\item The integrations needed for the computation of $\Sigma$ 
  are performed with the help of the self-written {\tt C++} program
  {\tt Xsection}.

  The SM part entering our numerical analysis is based on results obtained in
  Ref.~\cite{Pak:2011hs}. It is used to extract the quantities
  $\Sigma^{(i)}$ in Eq.~(\ref{eq::sigma_t}) which are necessary to construct
  the top sector result up to NNLO.

\item In the {\tt C++} program we have also included the computation of
  $\sigma_{tb}^{\rm SQCD}(\mu_s)|_{\rm NLO}$ which is based on the formulae
  provided in Refs.~\cite{Harlander:2004tp,Degrassi:2010eu}.  The analytic
  expressions for the amplitudes which we implemented in our numerical
  program are taken from Ref.~\cite{Degrassi:2010eu} for the bottom and from
  Ref.~\cite{Harlander:2004tp} for the top sector which provides the SUSY QCD
  corrections up to NLO. Thus the exact bottom quark mass dependence is taken
  into account as well as the top quark, gluino and squark dependence in the
  effective-theory approach, i.e. in the formal limit $M_h\to0$. For Higgs
  boson masses considered here this approximation approaches the exact result
  to high
  accuracy~\cite{Harlander:2009mq,Pak:2009dg,Harlander:2010my,Pak:2011hs}.  We
  have cross checked our implementation against the results implemented
  in~\cite{Harlander:2010wr} in certain limits.

\item We use LO, NLO and NNLO PDFs by the MSTW2008
  collaboration~\cite{Martin:2009iq} in order to obtain the respective
  predictions for $\sigma^{\rm SQCD}$ in
  Eq.~(\ref{eq::sigma}).  The choice of the PDFs determines the values of
  $\alpha_s(M_Z)\equiv\alpha_s^{(5)}(M_Z)$:
  \begin{eqnarray}
    \alpha_s^{\rm LO}(M_Z) &=& 0.1394\,,~
    \alpha_s^{\rm NLO}(M_Z) = 0.1202\,,~
    \alpha_s^{\rm NNLO}(M_Z) = 0.1171\,.
  \end{eqnarray}
  The appropriate $\overline{\rm MS}$ beta function then
  determines $\alpha_s(\mu_R)$ that enters the formulae.

\item The procedure for the calculation of the best three-loop approximation
  of $C_1$ within SQCD as described in Subsection~\ref{sub::appr} is
  implemented in {\tt Mathematica}.

\item The electroweak correction which enter the final result in factorized
  form (cf. Eq.~(\ref{eq::sigma})) are taken from
  Ref.~\cite{Actis:2008ug}.

\item
  If not stated otherwise the following default values are used:
  \begin{align}
    \mu_s = \frac{M_h}{2}\,,&&
    \mu_h = M_t\,,&&
    M_t   = 173.3~\mbox{GeV}\,.
  \end{align}

\end{itemize}


\subsection{Prediction to NNLO}

In order to discuss the numerical effects we start with the $m_h^{\rm max}$
scenario introduced in Ref.~\cite{Carena:2002qg}. The low-energy parameters
are given by
\begin{align}
  &&A_t=A_b=A_\tau = \sqrt{6}\,\msusy+\frac{\muSUSY}{\tan\beta}\,,
  &&\muSUSY=200~\mbox{GeV}\,,
  &&\msusy=1000~\mbox{GeV}\,
  \nonumber\\
  &&M_1 = 5 s_W^2 / (3 c_W^2) M_2\,,
  &&M_2 = 200~\mbox{GeV}\,,
  &&M_3 = 800~\mbox{GeV}\,,
  \label{eq::parameters2}
\end{align}
where the mass of the pseudo-scalar Higgs boson $M_A$ and the ratio of the
vacuum expectation values of the two Higgs doublets $\tan\beta$ is varied. In
Eq.~(\ref{eq::parameters2}) $A_t$ is the trilinear coupling, $\muSUSY$
is the Higgs-Higgsino bilinear coupling from the super potential, $m_{\rm
  susy}$ is the common mass value for all squarks, $M_1$, $M_2$ and $M_3$ are
the gaugino mass parameters and $s_W$ and $c_W$ are the sine and cosine of the
weak mixing angle.  The parameters in Eq.~(\ref{eq::parameters2}) are defined
at the scale $\mu_R=\sqrt{\msusy^2+M_t^2}$.

\begin{figure}[t]
  \centering
  \begin{tabular}{cc}
    \includegraphics[width=.48\linewidth]{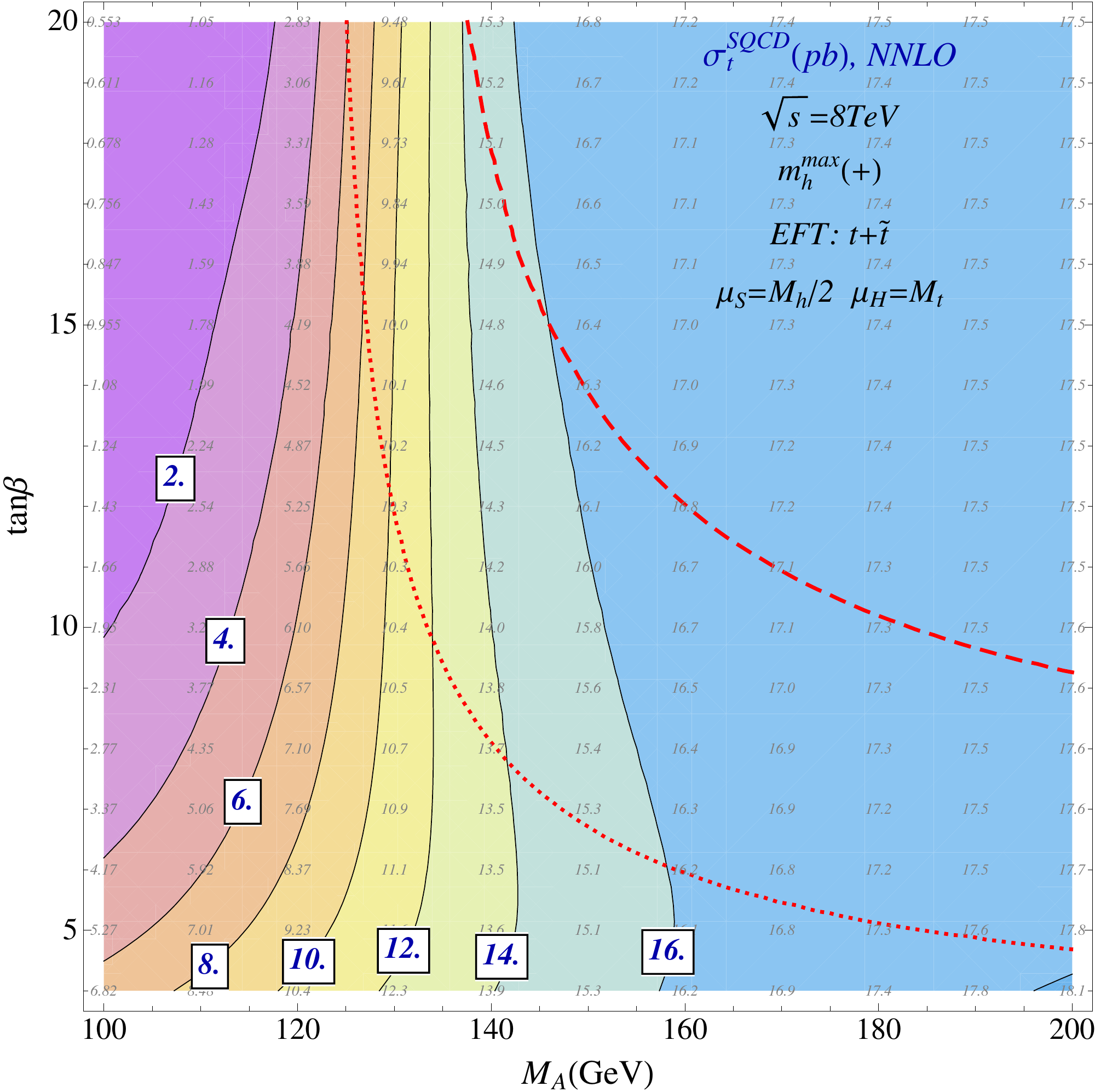}
    &
    \includegraphics[width=.48\linewidth]{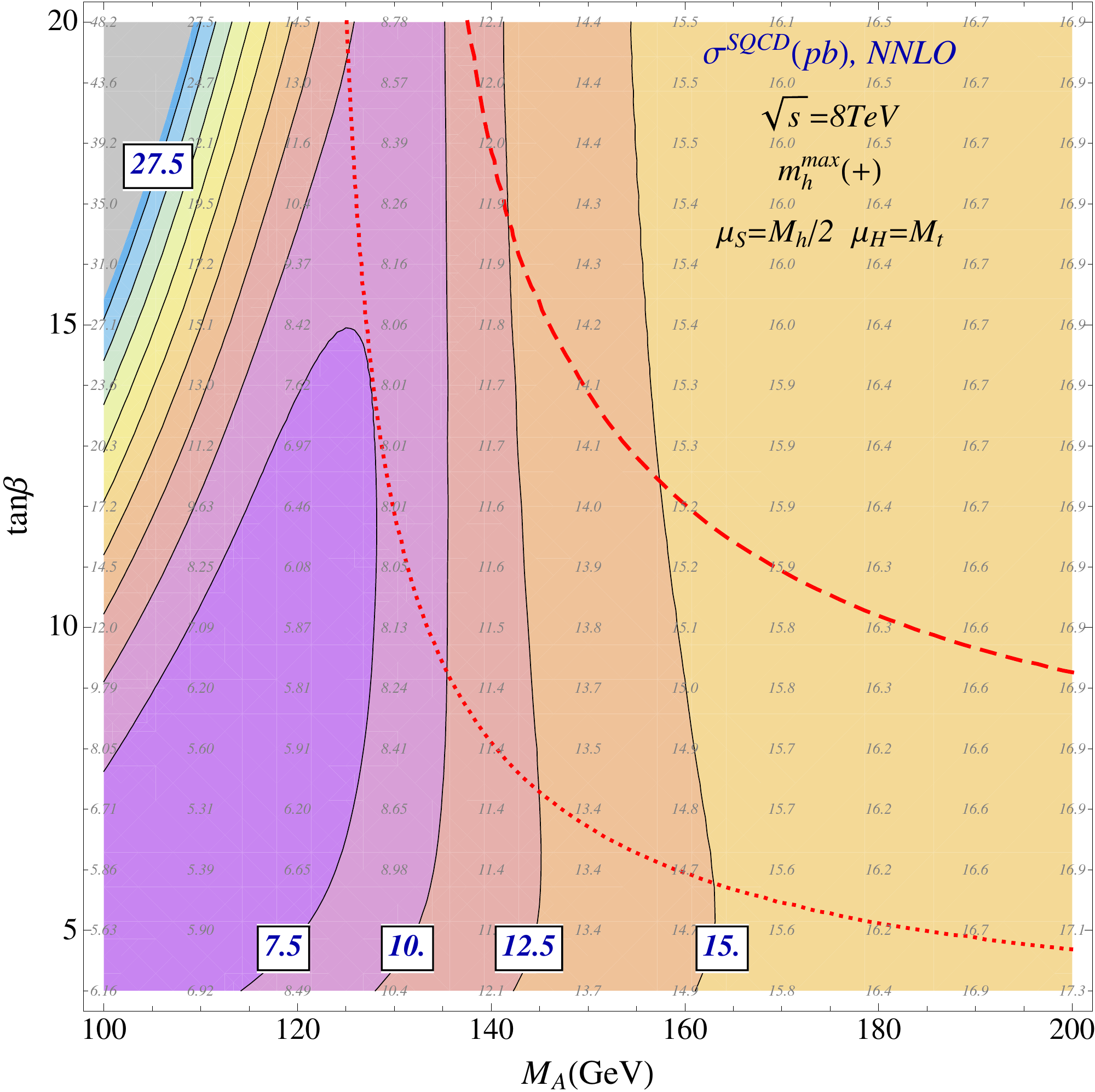}
    \\
    (a) & (b)
  \end{tabular}
  \caption[]{\label{fig::sig_mhmax_full}(a) $\sigma_t^{\rm SQCD}(\mu_s,\mu_h)$
    and (b) $\sigma_{\rm full}$ as a function of $M_A$ and $\tan\beta$ in the 
    $m_h^{\rm max}$ scenario.}
\end{figure}

In Fig.~\ref{fig::sig_mhmax_full} we discuss the dependence of the cross
section on $\tan\beta$ and $M_A$, the remaining parameters are fixed to the
values of Eq.~(\ref{eq::parameters2}).  The NNLO prediction for the top sector
contribution $\sigma_t^{\rm SQCD}(\mu_s,\mu_h)$ is shown in
Fig.~\ref{fig::sig_mhmax_full}(a) where the numbers on the contour
lines\footnote{The small gray numbers in the background provide the cross
  section in pb for various values of $\tan\beta$ and $M_A$. In order not to
  overload the plot it was chosen to keep these numbers small which means that
  they probably can only be read after magnifying the plots on the
  screen. Nevertheless we prefer to provide this information.}  correspond to
the cross section in pb.\footnote{Note that $\sigma_t^{\rm SQCD}$ contains
  also contributions where the Higgs boson couples to squarks corresponding to
  light quarks. Since these contributions are numerically small we write in
  the legend only ``$t+\tilde{t}$''.}  $M_A$ is varied between 100~GeV and
200~GeV and we choose $\tan\beta\le 20$.  The (red) hyperbola-like lines
indicate Higgs boson masses of 123~GeV (lower) and 129~GeV, respectively. Thus
the area in between is in accordance with the result reported by
ATLAS~\cite{ATLAS-Higgs} and CMS~\cite{CMS-Higgs}.
The NNLO cross section has been obtained with the help of hierarchy (h1)
which is appropriate since the SUSY mass spectrum is almost degenerate.

For comparison we show in Fig.~\ref{fig::sig_mhmax_full}(b) the complete
result including also bottom quark effects and the electroweak corrections
according to Eq.~(\ref{eq::sigma}).  One observes a significant enhancement of
the cross section for small values of $M_A$ and large $\tan\beta$ (i.e. in the
upper left corner).  In the remaining parts of the plot only a reduction of a
few percent is observed which comes basically from the destructive
interference of the top
and bottom contribution. Let us mention that a large part of the parameter
space shown in Fig.~\ref{fig::sig_mhmax_full} is already excluded
experimentally which will be discussed below.  Basically only the lower right
part of the plot is experimentally still allowed. Note that this region is
dominated by the contribution from the top sector and the bottom contribution
only plays a minor role.

\begin{figure}[t]
  \centering
  \begin{tabular}{cc}
    \includegraphics[width=.48\linewidth]{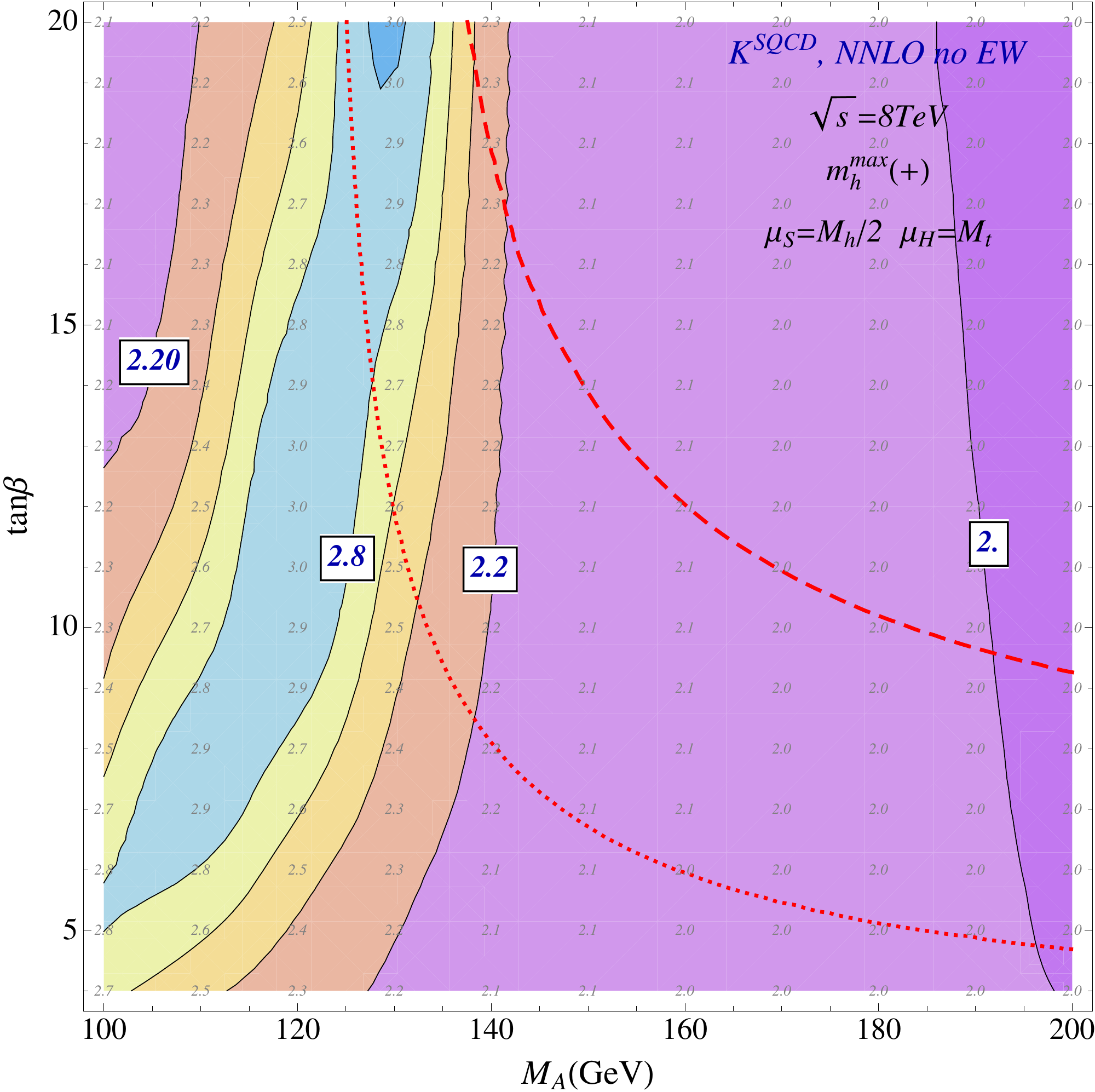}
    &
    \includegraphics[width=.48\linewidth]{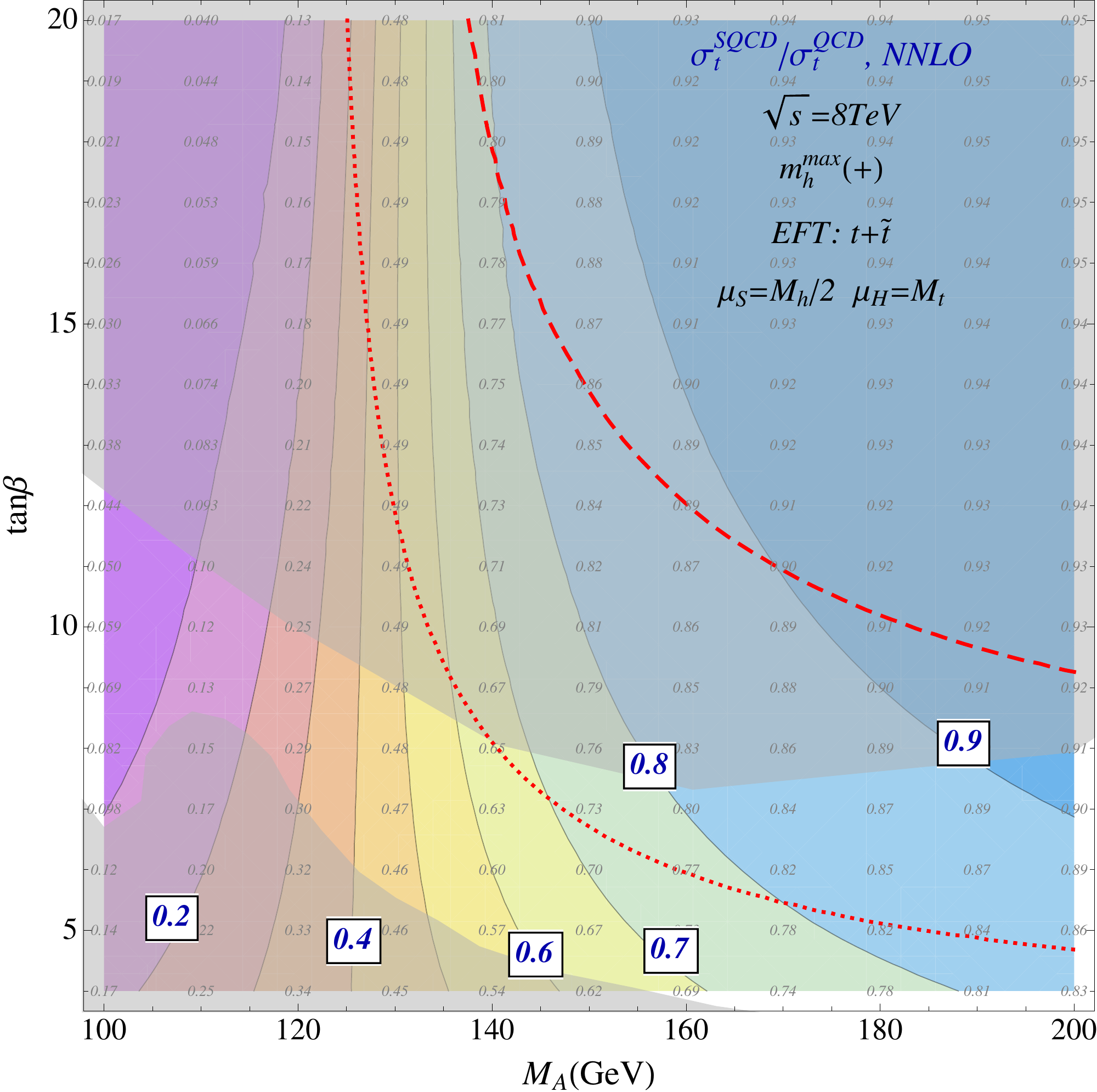}
    \\
    (a) & (b)
  \end{tabular}
  \caption[]{\label{fig::k_mhmax_nnlo}(a) NNLO $K$ factor corresponding to the
    prediction in Fig.~\ref{fig::sig_mhmax_full}(b).
    Panel (b) shows the ratio of the MSSM and the SM NNLO prediction from
    the top quark/top squark sector.}
\end{figure}

The effect of higher order corrections is conveniently parametrized in terms of
the $K$ factor which we define through
\begin{eqnarray}
  K &=& \frac{\sigma^{\rm HO}}{\sigma^{\rm LO}}
  \,.
  \label{eq::K}
\end{eqnarray}
In Fig.~\ref{fig::k_mhmax_nnlo}(a) we show the NNLO (i.e. HO=NNLO in
Eq.~(\ref{eq::K})) $K$ factor for the same parameter space as in
Fig.~\ref{fig::sig_mhmax_full}. In the experimentally allowed region it
amounts to about 100\% where about 70\% arise from the NLO corrections.  It is
interesting to compare the MSSM prediction to the one of the SM which is done
in Fig.~\ref{fig::k_mhmax_nnlo}(b) where the ratio of the NNLO predictions in
the two models is plotted.  One observes that the supersymmetric corrections
reduce the cross section by typically 60\% to 90\% in the parameter space
where the Higgs boson mass is between 123~GeV and 129~GeV.  Note, however,
that for $M_A\gsim 140$~GeV this effect basically results from a modification
of the Higgs couplings to fermions and thus is already present at LO. In fact
for this region the $K$ factors in the SM and MSSM are very similar for the
$m_h^{\rm max}$ scenario.  In Fig.~\ref{fig::k_mhmax_nnlo}(b) we also show the
regions excluded by searches performed at {\tt LEP}~\cite{Schael:2006cr} and
{\tt CMS}~\cite{Chatrchyan:2012vp} which is indicated by the gray
area.\footnote{The data points have been extracted from Fig.~3 of
  Ref.~\cite{Chatrchyan:2012vp} using the program {\tt
    EasyNData}~\cite{Uwer:2007rs}.}

The reason for the small deviations between the SM and MSSM in the $m_h^{\rm
  max}$ scenario  is the heavy supersymmetric mass spectrum resulting from the
input parameters in Eq.~(\ref{eq::parameters2}). Actually, the typical squark
mass is 1000~GeV with only small differences between the individual
flavours and the gluino mass takes the value $\mgluino=860$~GeV.

In the following we slightly modify the parameters in such a way that one 
of the top squarks becomes light whereas the other remains heavy. Thus, on one
hand one obtains sizeable supersymmetric contributions to the Higgs boson
production cross section and at the same time a Higgs boson mass which is
sufficiently heavy. To date such a scenario has not been excluded by
experimental searches.

In our modified $m_h^{\rm max}$ scenario
the default values of the low-energy parameters are given by
\begin{align}
  &&A_b=A_\tau=2469.48~\mbox{GeV}\,,
  &&A_t=1500~\mbox{GeV}\,,
  &&\muSUSY=200~\mbox{GeV}\,,
  \nonumber\\
  &&M_1 = 5 s_W^2 / (3 c_W^2) M_2\,,
  &&M_2 = 200~\mbox{GeV}\,,
  &&M_3 = 800~\mbox{GeV}\,,
  \nonumber\\
  &&m_{\rm susy}=1000~\mbox{GeV}\,,
  &&\tilde{m}_t = 400~\mbox{GeV}\,,
  \nonumber\\
  &&M_A=1000~\mbox{GeV}\,,
  &&\tan\beta = 20\,,
  \label{eq::parameters3}
\end{align}
where $\tilde{m}_t$ represents the singlet soft SUSY breaking parameter of the
right-handed top squark (i.e. {\tt EXTPAR 46} in the ``Les Houches 
Accord'' notation~\cite{Skands:2003cj,Allanach:2008qq}).

\begin{figure}[t]
  \centering
  \begin{tabular}{c}
    \includegraphics[width=.8\linewidth]{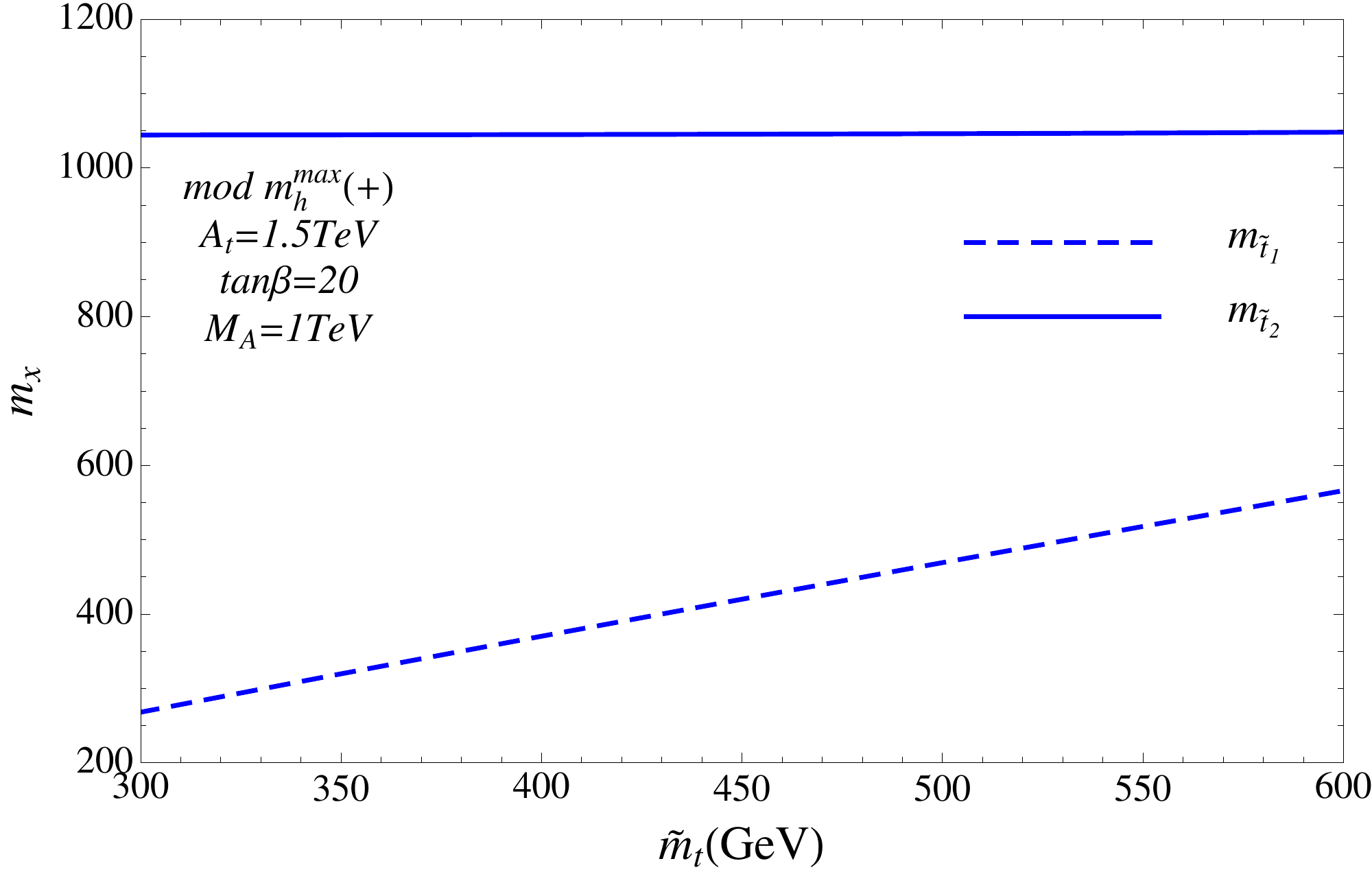}
    \\ (a) \\
    \includegraphics[width=.8\linewidth]{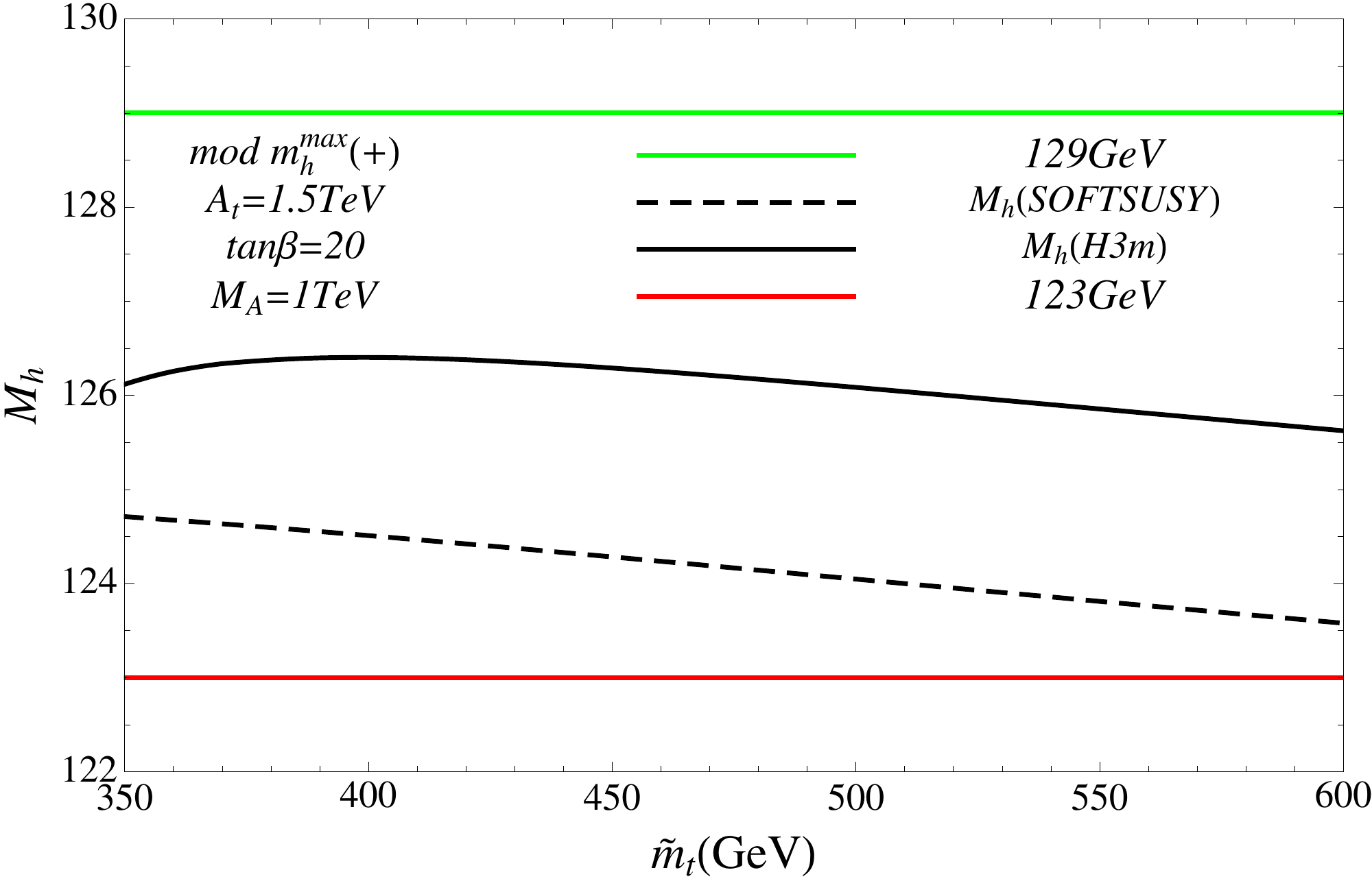}
    \\
    (b)
  \end{tabular}
  \caption[]{\label{fig::spect}(a) Top squark masses and (b) lightest 
    Higgs
    boson mass as a function of $\tilde{m}_t$ for the parameters
  given in Eq.~(\ref{eq::parameters2}).}
\end{figure}

In Fig.~\ref{fig::spect}(a) we show the top squark masses as a function of
$\tilde{m}_t$. The variation of this parameter essentially changes $\mstopone$
whereas $\mstoptwo$ remains almost constant at a large value of about
1050~GeV. For large values of $\tilde{m}_t$ one obtains a heavy spectrum and
thus the supersymmetric corrections to the Higgs boson production become
small. On the other hand, for a small value of $\tilde{m}_t$ and thus a large
splitting in the top squark sector the numerical effect can be large as we
will discuss below. 
The hierarchy in the top squark sector suggests to use (h2) 
for evaluating the three-loop corrections to $C_1$.

It is interesting to have a closer look at the numerical values of the
three-loop coefficient $c_1^{(2)}$.  For $\tilde{m}_t=400$~GeV we
have\footnote{The numbers in the brackets indicate the uncertainty according
  to our approximation procedure.} $c_1^{(2)}(\mu=M_t)\approx -0.58(1)$ and
$c_1^{(2)}(\mu=M_h/2\approx 63~\mbox{GeV})\approx -14.66(1)$ and for
$\tilde{m}_t=600$~GeV one obtains $c_1^{(2)}(\mu=M_t)\approx 5.70(1)$ and
$c_1^{(2)}(\mu=M_h/2\approx 63~\mbox{GeV})\approx -1.29(23)$.  One observes a
quite strong dependence of $c_1^{(2)}$ on the renormalization scale which
demonstrates the importance of the three-loop calculation.  Note that the
renormalization scale dependence is compensated by the other quantities
entering formula~(\ref{eq::sigma_t}) for the cross section.

In order to get a feeling about the physical masses we list in the following
the $\overline{\rm DR}$ mass values for the squark and gluino masses. For
$\mu_R=M_t$ and $\tilde{m}_t=400$~GeV one obtains from the parameters provided
by {\tt \softsusy} the following values
\begin{align}
  & \mstopone = 370~\mbox{GeV}\,, & \mstoptwo = 1045~\mbox{GeV}\,, & 
  \nonumber\\
  & \msquark = 1042~\mbox{GeV}\,, & \mgluino = 860~\mbox{GeV}\,, & 
\end{align}
where $\msquark$ corresponds to the average of
$m_{\tilde{u}}$, $m_{\tilde{d}}$, $m_{\tilde{s}}$ and $m_{\tilde{c}}$.

The lightest MSSM Higgs boson mass $M_h$ as predicted from the parameters in
Eq.~(\ref{eq::parameters2}) is shown in Fig.~\ref{fig::spect}(b) where
$\tilde{m}_t$ is varied in the same range as before. In this figure we compare
$M_h$ as computed by {\tt \softsusy} (dashed lines) to the one obtained by {\tt
  H3m}~\cite{Harlander:2008ju,Kant:2010tf} (solid line). Note that the latter,
which includes three-loop corrections, is more than 1~GeV above the former.
As one can see, in the whole range of $\tilde{m}_t$ the prediction remains
within the horizontal lines which mark 123~GeV and 129~GeV, respectively, the
limits motivated by the recent results reported by ATLAS~\cite{ATLAS-Higgs} and
CMS~\cite{CMS-Higgs}.

\begin{figure}[t]
  \centering
  \begin{tabular}{c}
    \includegraphics[width=.8\linewidth]{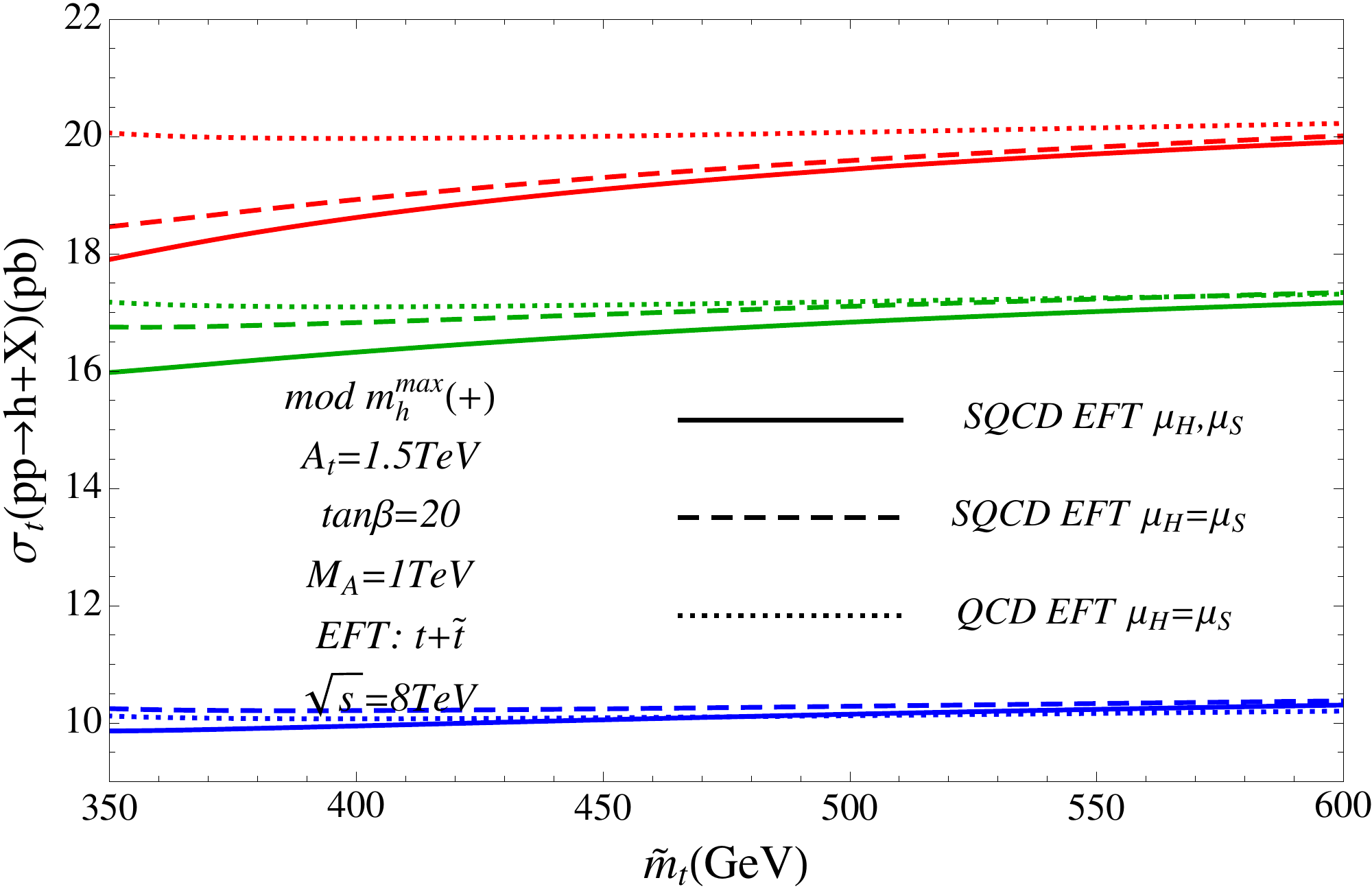}
  \end{tabular}
  \caption[]{\label{fig::sig_mtil}The cross section $\sigma_t^{\rm
      SQCD}(\mu_s,\mu_h)$ as a function of $\tilde{m}_t$ using the parameters
    in Eq.~(\ref{eq::parameters2}) to LO (bottom), NLO (middle) and NNLO
    (top).  The dotted line corresponds to the SM and the dashed and solid
    lines to the MSSM.}
\end{figure}

In Fig.~\ref{fig::sig_mtil} we show $\sigma_t^{\rm SQCD}(\mu_s,\mu_h)$ as a
function of $\tilde{m}_t$ at LO, NLO and NNLO (from bottom to top). For each
order three curves are shown where the dotted curve corresponds to the SM. The
SQCD corrections are included in the dashed and solid line where for the
former the soft and hard scale have been identified with $M_h/2$ and for the
latter $\mu_s=M_h/2$ and $\mu_h=M_t$ has been chosen. As already mentioned
above, the difference between the SM and MSSM corrections is small for large
values of $\tilde{m}_t$, for smaller values, however, a sizeable effect is
visible. For example, for $\tilde{m}_t=400$~GeV a reduction of the cross
section of about 5\% is observed.

The difference between the dashed and solid line quantifies the effect of the
resummation discussed in Subsection~\ref{sub::form}. It is negligible for
large $\tilde{m}_t$, however, for smaller values it can be as large as
2-3\%. It is interesting to note that supersymmetric three-loop
corrections to $C_1$ computed in this paper play an important role. In
case 
the three-loop coefficient is identified with the SM one
a reduction of only 3\% and not 5\% is observed.

\begin{figure}[t]
  \centering
  \begin{tabular}{c}
    \includegraphics[width=.8\linewidth]{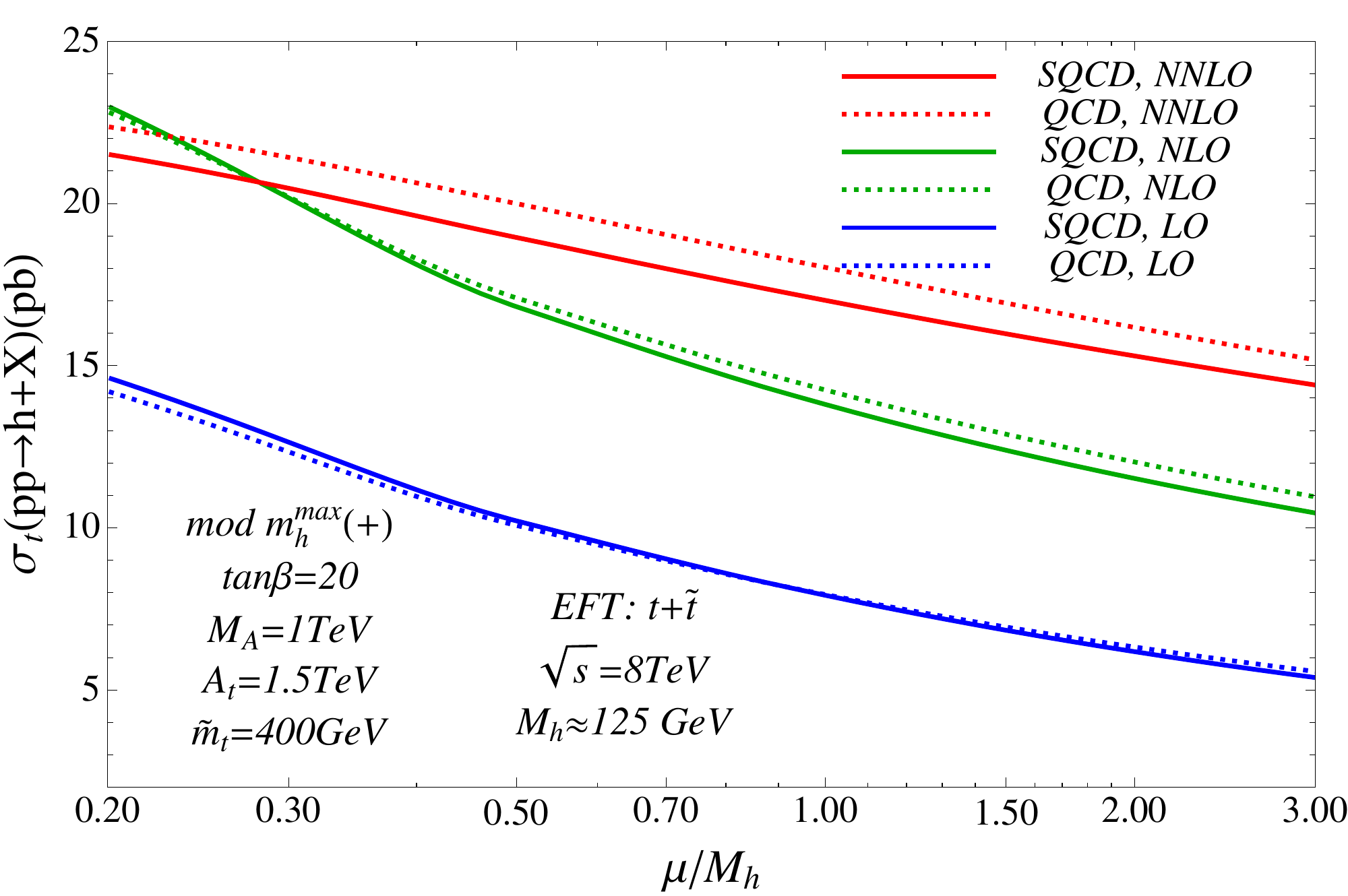}
  \end{tabular}
  \caption[]{\label{fig::sig_mu}Renormalization scale dependence of
    the cross section $\sigma_t^{\rm SQCD}(\mu_s=\mu_h)$.}
\end{figure}

In Fig.~\ref{fig::sig_mu} we fix $\tilde{m}_t=400$~GeV and discuss the
renormalization scale dependence of the LO, NLO and NNLO result by varying
$\mu_h=\mu_s$ between 25~GeV and about 400~GeV.
The SM (dotted) and the MSSM (solid lines) curves show a very
similar behaviour.  Close to the central scale $\mu_s=M_h/2$ one observes a
crossing of the NNLO and NLO curves, i.e., vanishing NNLO corrections. It is
at a slightly higher value for the MSSM which can be explained by the larger
masses in the spectrum entering the predictions.

\begin{figure}[t]
  \centering
  \begin{tabular}{c}
    \includegraphics[width=.8\linewidth]{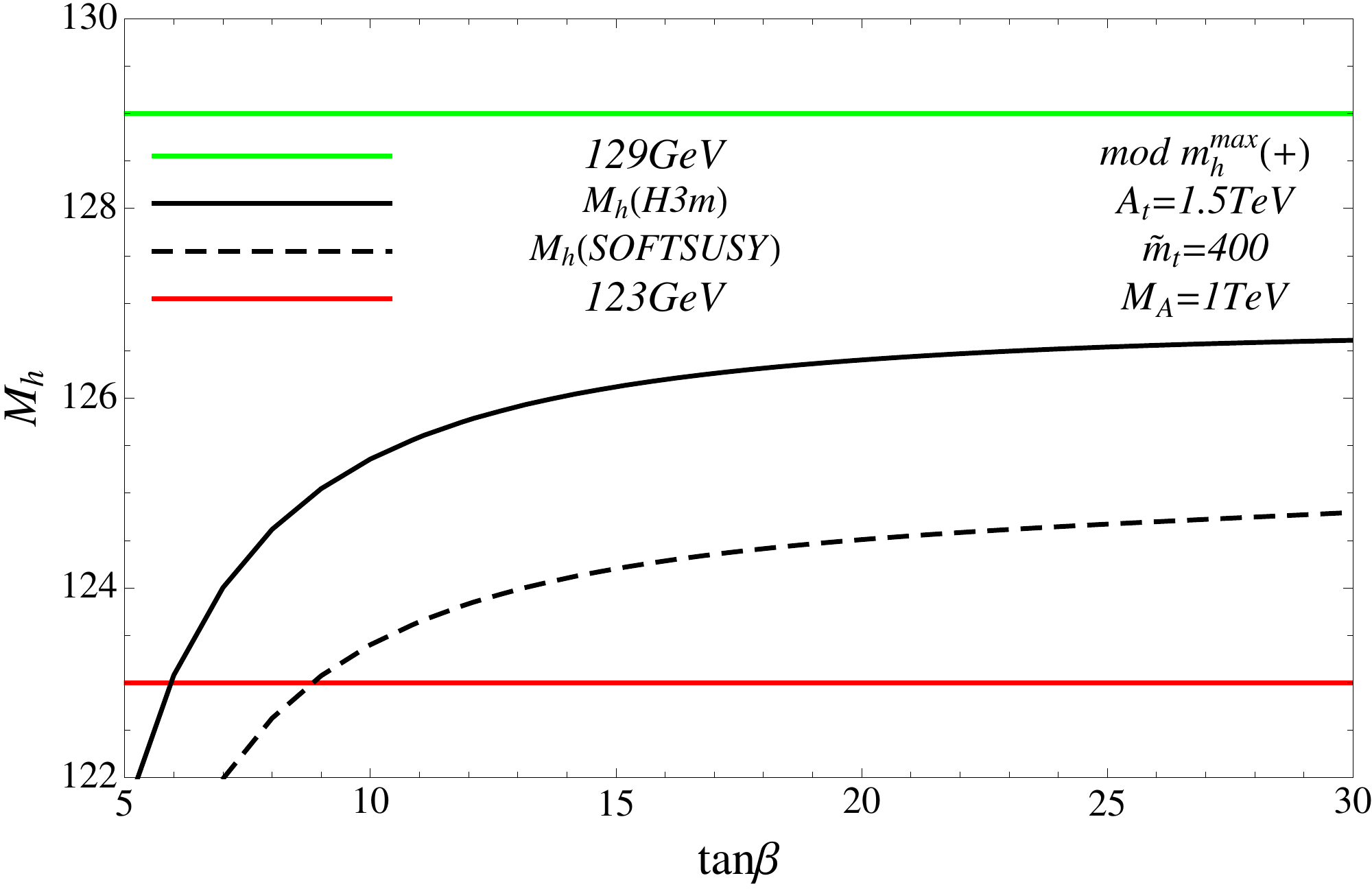}
  \end{tabular}
  \caption[]{\label{fig::spect_tb} Lightest Higgs
    boson mass as a function of $\tan\beta$ for the parameters
    given in Eq.~(\ref{eq::parameters3}).}
\end{figure}

In a next step we study the dependence on $\tan\beta$ for fixed
$\tilde{m}_t=400$~GeV. One observes top squark masses which are almost
constant and are given by $\mstopone\approx 370$~GeV and
$\mstoptwo\approx 1040$~GeV. On the other hand there is a strong dependence of
$M_h$ on $\tan\beta$ which is shown in Fig.~\ref{fig::spect_tb} for the 
result from {\tt \softsusy} and {\tt H3m}. As one can see, for $\tan\beta\le 6$
the predicted Higgs boson mass is below 123~GeV using {\tt H3m}; on the basis
of {\tt \softsusy} this boundary is reached already for $\tan\beta\approx 9$.

\begin{figure}[t]
  \centering
  \begin{tabular}{c}
    \includegraphics[width=.8\linewidth]{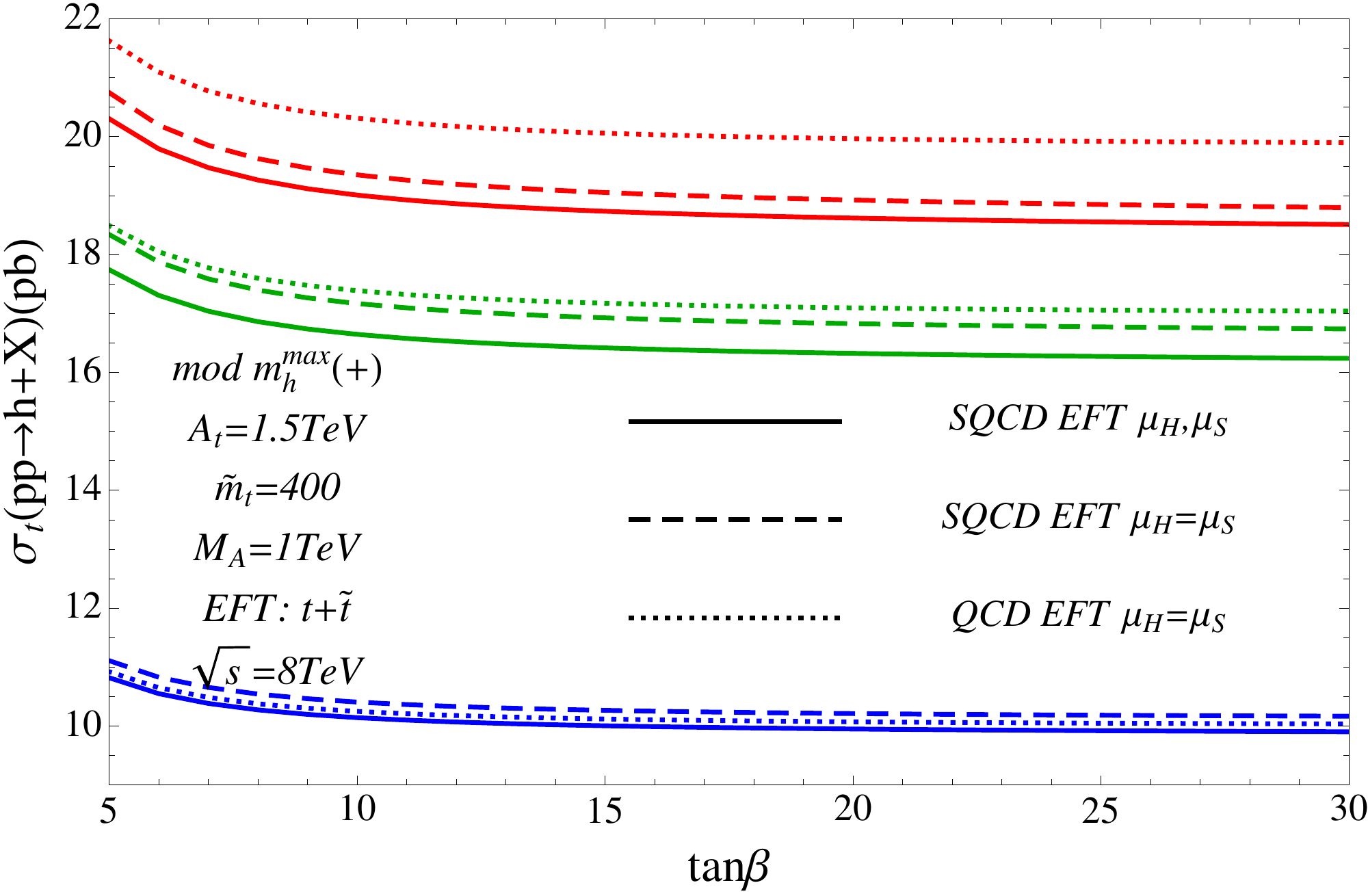}
  \end{tabular}
  \caption[]{\label{fig::sig_tb}The cross section $\sigma_t^{\rm
      SQCD}(\mu_s,\mu_h)$ as a function of $\tan\beta$ using the parameters
    in Eq.~(\ref{eq::parameters3}) to LO (bottom), NLO (middle) and NNLO
    (top).  The dotted line corresponds to the SM and the dashed and solid
    lines to the MSSM.}
\end{figure}

In Fig.~\ref{fig::sig_tb} we show $\sigma_t^{\rm SQCD}(\mu_s,\mu_h)$ as a
function of $\tan\beta$ for $5\le\tan\beta\le 30$ at LO, NLO and NNLO (from
bottom to top) choosing $A_t=1500$~GeV and $\tilde{m}_t=400$~GeV.  The SM
result is again shown as dotted line. The MSSM predictions are both shown for
$\mu_h=\mu_s=M_h/2$ (dashed) and $\mu_h=M_t$ and $\mu_s=M_h/2$ (solid line).
As compared to the SM prediction the MSSM results are again reduced by a few
percent. Note that this effect increases when going from LO to NLO and finally
to NNLO where a difference of about 5\% is observed. It is also interesting
to mention that the proper choice of scales is not negligible: the results for
$\mu_h=\mu_s$ are in general a few percent above the ones with
$\mu_h\not=\mu_s$.

\begin{figure}[t]
  \centering
  \begin{tabular}{c}
    \includegraphics[width=.8\linewidth]{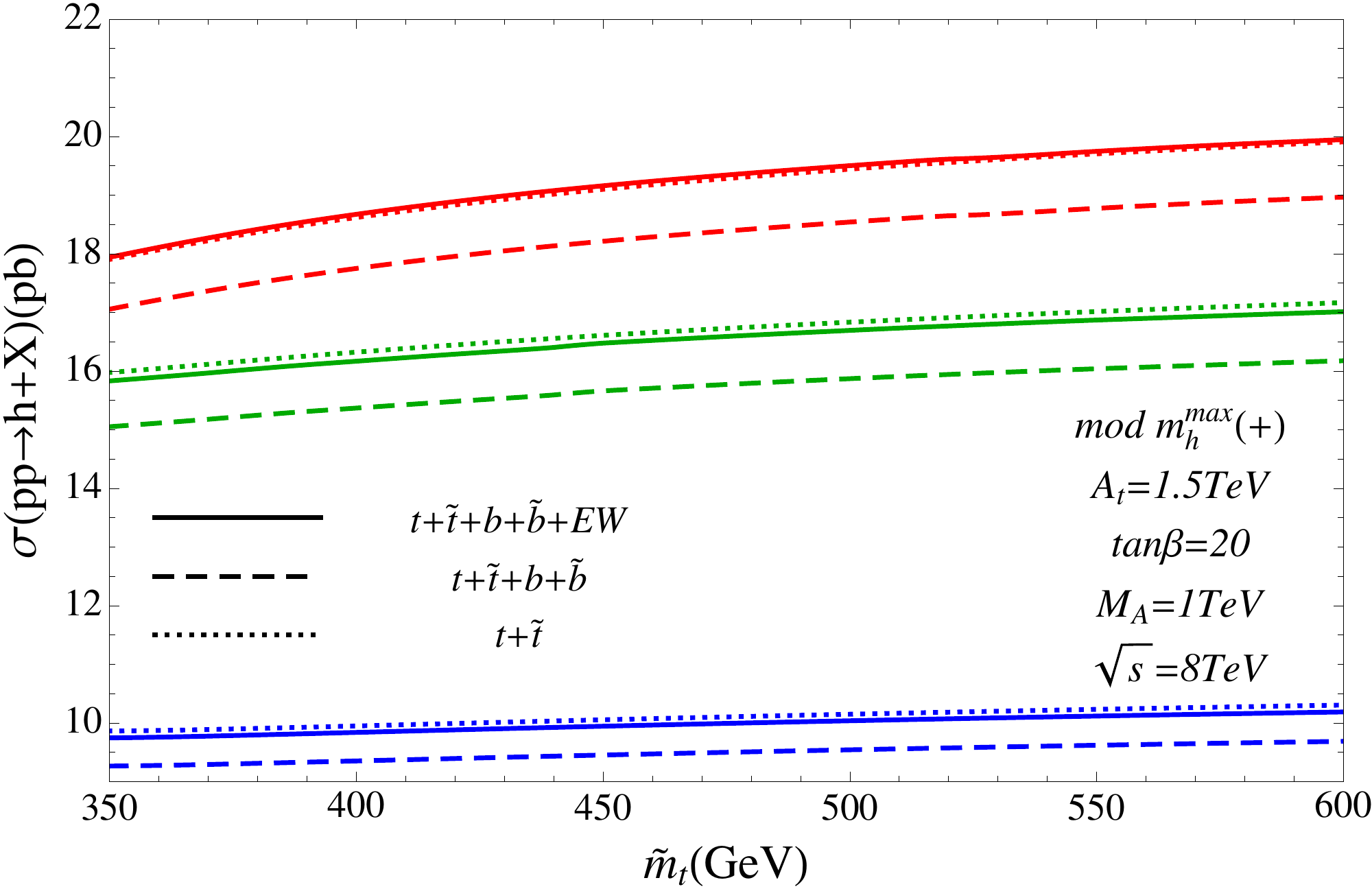}
    \\ (a) \\
    \includegraphics[width=.8\linewidth]{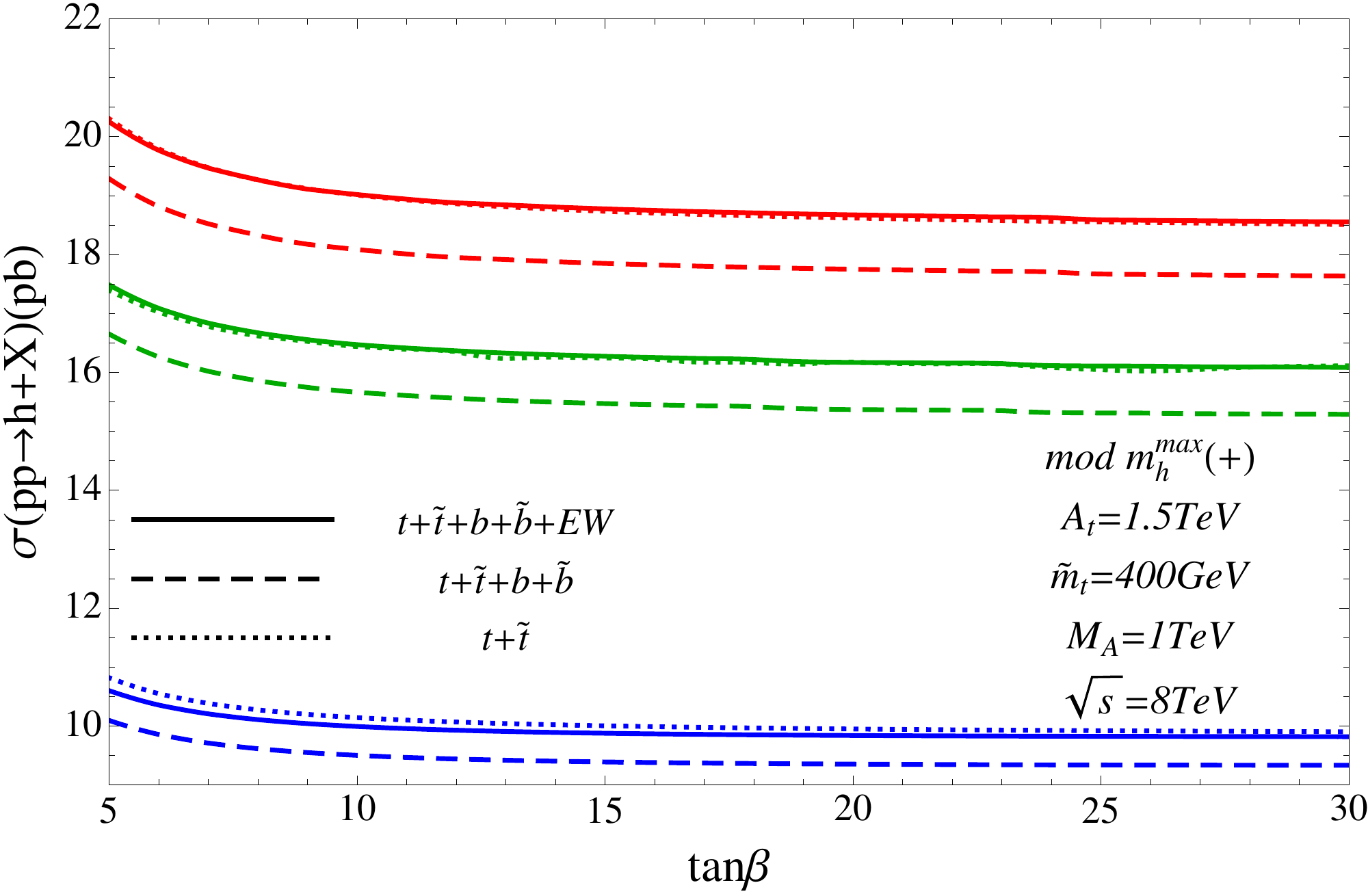}
    \\ (b) 
  \end{tabular}
  \caption[]{\label{fig::sig_H_mtil_tb}The cross section $\sigma^{\rm
      SQCD}(pp\to h+X)$ 
    as a function of (a) $\tilde{m}_t$ and (b) $\tan\beta$ using the parameters
    in Eq.~(\ref{eq::parameters3}).}
\end{figure}

Let us finally present results for $\sigma(pp\to h+X)$ which include in
addition bottom quark contributions up to NLO and furthermore also electroweak
corrections.  In Fig.~\ref{fig::sig_H_mtil_tb}(a) and~(b) the dependence on
$\tilde{m}_t$ and $\tan\beta$ is plotted, respectively, where from bottom to
top the LO, NLO and NNLO results are shown.  We refrain from showing the
bottom quark contributions separately since they are about two orders of
magnitude smaller than the top contributions. However, the interference terms
have a visible effect which can be seen by the difference between the dashed
and dotted curves since in the latter the quantities $\sigma_{tb}^{\rm
  SQCD}(\mu_s)|_{\rm NLO}$ and $\sigma_t^{\rm SQCD}(\mu_s)|_{\rm NLO} $ in
Eq.~(\ref{eq::sigma}) are set to zero. One observes a reduction of about 5\%
after including bottom quark effects --- basically independent of $\mtildeuR$
and $\tan\beta$. Thus even for $\tan\beta=20$ the bottom quark effects are
small for the considered scenarios and hence a NNLO
calculation for $m_b\not=0$ is not mandatory.  The reduction from top/bottom
interference is to a large extend compensated by the electroweak corrections
taken into account in a multiplicative way (cf. Eq.~(\ref{eq::sigma})) as can
be seen by solid line which includes all contributions of
Eq.~(\ref{eq::sigma}).



\section{\label{sec::concl}Conclusions}

We have computed the three-loop corrections to the matching coefficient of the
effective Higgs-gluon coupling $C_1$ originating from supersymmetric QCD. The
three-loop integrals contain several mass scales and thus an exact calculation
on the basis of the techniques available these days is not possible. We have
developed an approximation method based on expansions in mass hierarchies and
mass differences which selects the best parametrization for the expansion
parameters. The described procedure is not bound to the problem at hand but
can certainly also be applied to other processes.

The three-loop result for $C_1$ is used in order to evaluate the Higgs boson
production cross section at LHC to NNLO. Whereas the three-loop expressions
are only available for vanishing light quark masses we include up to NLO the
contributions from the top and bottom sector allowing for a non-vanishing
bottom quark mass. Numerical results for the total production cross section
are presented for a center-of-mass energy of $\sqrt{s}=8$~TeV for two MSSM
scenarios which predict a light supersymmetric Higgs boson mass in accordance
with the recent results from ATLAS and CMS. We study both the dependence on
the renormalization scale, $\tan\beta$ and on the $\mtildeuR$, the singlet
soft SUSY breaking mass parameter of the right-handed top squark.
For the case where all supersymmetric masses are above 1~TeV the
supersymmetric corrections are small. However, in case one of the top squarks
becomes of the order of a few hundred GeV the supersymmetric corrections to
the matching coefficient $C_1$ are not negligible any more and 
a sizeable effect on the production cross section is noticeable.
For the scenarios considered in this paper a reduction of the NNLO prediction
of the cross section by a few percent is observed. We also want to stress that
visible effects result from separating the soft and hard scale in $C_1$ which 
automatically resums potentially large logarithms involving the Higgs boson
mass.

In order to perform the numerical evaluation a flexible tool has been
developed which allows a convenient handling of the spectrum generators in
combination with {\tt Mathematica} and {\tt C++} routines evaluating the 
cross sections. In a comfortable way it is possible to vary the
renormalization scale or other parameters entering the prediction.



\section*{Acknowledgements}

We thank Pietro Slavich for communications in connection to
Refs.~\cite{Degrassi:2008zj,Degrassi:2010eu} and for providing us with the {\tt
  Fortran} routines for the virtual corrections. We thank Robert Harlander for
discussions on Ref.~\cite{Harlander:2010wr} and carefully reading the
manuscript,  Thomas Hermann and Luminita
Mihaila for providing computer-readable  expressions for the two-loop
counterterms and Philipp Kant for the support in connection to {\tt H3m}.  This
work was supported by the DFG through the SFB/TR~9 ``Computational Particle
Physics''.




\end{document}